\documentclass[range]{ar2e}
                           
\usepackage{graphicx}
\usepackage{bm}
\usepackage{epsfig}
\usepackage{epstopdf}
\usepackage{bm}
\usepackage{epsfig}
\usepackage{graphics}
\usepackage{xspace}
\usepackage{pstricks}
\usepackage{subfigure}
\usepackage{amssymb}
\usepackage{amsmath,wasysym}
\usepackage{verbatim}

\def\OMIT#1{}

\def\msbar{$\overline{\rm MS}$}

\def\to{\rightarrow}

\def\ovl{\overline}

\def\bmat{\begin{pmatrix}}
\def\emat{\end{pmatrix}}
\def\ph{\phantom}

\def\sstw{{\sin^2\theta_W}}

\newcommand{\ba}{\begin{array}}
\newcommand{\ea}{\end{array}}

\newcommand{\nn}{\nonumber}

\newcommand{\bea}{\begin{eqnarray}}
\newcommand{\eea}{\end{eqnarray}}

\newcommand{\be}{\begin{equation}}
\newcommand{\ee}{\end{equation}}

\newcommand{\bfl}{\begin{flushleft}}
\newcommand{\bfr}{\begin{flushright}}
\newcommand{\efl}{\end{flushleft}}
\newcommand{\efr}{\end{flushright}}

\newcommand{\beq}{\begin{equation}}
\newcommand{\eeq}{\end{equation}}
\newcommand{\beqa}{\begin{eqnarray}}
\newcommand{\eeqa}{\end{eqnarray}}

\newcommand{\lsim}{\mathrel{\lower4pt\hbox{$\sim$}}
\hskip-12.5pt\raise1.6pt\hbox{$<$}\;}
\newcommand{\gsim}{\mathrel{\lower4pt\hbox{$\sim$}}
\hskip-11.5pt\raise1.6pt\hbox{$>$}\;}

\newcounter{saveeqn}

\begin{document}

\input epsf.tex    
\input epsf.def   

\input psfig.sty

\jname{Annu. Rev. Nucl. Part. Sci.}
\jyear{2013}
\jvol{1}
\ARinfo{1056-8700/97/0610-00}

\title{Weak Polarized Electron Scattering}

\markboth{Erler, Horowitz, Mantry, \& Souder}{Weak Polarized Electron Scattering}

\author{Jens Erler
\affiliation{PRISMA Cluster of Excellence \& Mainz Institute for Theoretical Physics, 
Johannes Gutenberg University, D-55099 Mainz, Germany\\
Permanent address: Departamento de F\'isica Te\'orica, Instituto de F\'isica, 
Universidad Nacional Aut\'onoma de M\'exico, M\'exico D.F. 04510, M\'exico}
Charles J. Horowitz
\affiliation{Department of Physics and Nuclear Theory Center, Indiana University, Bloomington, IN 47405, USA}
Sonny Mantry
\affiliation{High Energy Division, Argonne National Laboratory,  Argonne, IL 60439\\
Department of Physics and Astronomy,
Northwestern University, \\ Evanston, IL 60208}
Paul A. Souder
\affiliation{Physics Department, Syracuse University, Syracuse, New York 13244}
}


\begin{abstract}

Scattering polarized electrons provides an important probe of the weak interactions.  
Precisely measuring the parity-violating left-right cross section asymmetry 
is the goal of a number of experiments recently completed or in progress.
The experiments are challenging, since $A_{LR}$  is small, typically
between 10$^{-4}$ and 10$^{-8}$.  By carefully choosing appropriate 
targets and kinematics, various pieces of the weak Lagrangian can be isolated,
providing a search for physics beyond the Standard Model.  
For other choices, unique features of the strong interaction are studied,
including the radius of the neutron density in heavy nuclei, charge symmetry violation,
and higher twist terms.  This article reviews the theory behind the experiments,
as well as the general techniques used in the experimental program.

\end{abstract}

\maketitle

\section{INTRODUCTION}\label{intro}
\subsection{Overview}

In 1978, SLAC experiment E122~\cite{Prescott:1978tm,Prescott:1979dh} published 
the observation of parity violation in the deep inelastic scattering of polarized electrons from deuterium.  
It settled the issue as to whether or not the then recently discovered weak 
neutral currents were parity-violating and led to the universal acceptance of the Standard Model (SM).
The experiment also demonstrated that parity violation in electron scattering (PVES) is a viable tool for
particle and nuclear physics.  

In the following 35 years, many new PVES experiments were performed or are planned at various
laboratories around the world, including SLAC, Mainz, MIT--Bates, and JLab.   
The goals of the new experiments included searching for non-zero strange elastic form factors~\cite{Armstrong:2012bi}, 
measuring the weak mixing angle, $\sin^2\theta_W$, at low energies~\cite{Kumar:2013yoa}, 
measuring a set of electroweak (EW) couplings, 
measuring the radius $R_n$ of the distribution of neutrons in heavy nuclei, 
searching for charge symmetry violation (CSV) at the quark level, 
and measuring higher-twist effects in deep inelastic scattering (DIS).  
The first two topics have been the subject of previous reviews~\cite{Armstrong:2012bi,Kumar:2013yoa}; 
here we will focus on the other topics.
 
The basic idea of PVES is to measure the parity-violating EW asymmetry,
\be 
A_{LR}=\frac{\sigma_L - \sigma_R}{\sigma_L + \sigma_R}\ ,
\ee
where $\sigma_L$ $(\sigma_R)$ is the cross section for the scattering of electrons with left (right) helicity.
The leading order effect arises from an interference between photon and $Z$ exchange, 
resulting in small asymmetries proportional to the four momentum transfer $Q^2$ and also to the EW couplings.
Values of $A_{LR}$ in the range from $10^{-4}$ to $10^{-8}$ can be measured with good accuracy. 
By selecting optimal kinematics and targets, the wide variety of physics mentioned above can be accessed.
  
A useful feature of $A_{LR}$ is that in the ratio a number of experimental uncertainties cancel, 
such as those due to target thickness and the solid angle.  
Some possible theoretical uncertainties, such as elastic form factors, also cancel.  
The kinematics of the various experiments are chosen either so that unknown hadronic effects cancel,
probing EW physics,  or so that the hadronic effects remaining are of interest and cannot be measured by other techniques.

\subsection{Electroweak Physics}
\label{intro3}
An efficient way to represent the sensitivities of different experiments and observables to
the underlying EW physics, including possible new physics beyond the SM,
is to use the language of a low-energy effective field theory in a way described in Section~\ref{sectheory}.
The parity-violating (PV) part of the neutral-current (NC) interactions of charged fermions with electrons is
\be\label{LeqNC}
{\cal L}_{\rm NC}^{\, e f} = {\ovl e \gamma^\mu\gamma^5 e \over 2 v^2} 
\left( \sum\limits_{q=u,d} g_{AV}^{\, e q} \ovl{q} \gamma_\mu q + 
{1\over 2}\, g_{AV}^{\, e e} \ovl{e} \gamma_\mu e \right) + 
{\ovl e \gamma^\mu e \over 2 v^2} \sum\limits_{q=u,d} g_{VA}^{\, e q} \ovl{q} \gamma_\mu\gamma^5 q\ ,
\ee
where $v = (\sqrt{2} G_F)^{-1/2} = 246.22$~GeV and $G_F$ is the Fermi constant.
The EW coefficients are real-valued and at the SM tree level given by
\be\label{gAVeud}
g_{AV}^{\, e u} = - {1\over 2} + {4\over 3} \sin^2\theta_W\ ,
\qquad\qquad
g_{AV}^{\, e d} = {1\over 2} - {2\over 3}  \sin^2\theta_W\ ,
\ee
\be\label{gVAeud}
g_{AV}^{\, e e} = g_{VA}^{\, e d} = - g_{VA}^{\, e u}  = {1\over 2} - 2 \sin^2\theta_W\ .
\ee
We now briefly review how these couplings are extracted experimentally.

Currently, the most precise determination of any combination derives 
from observations~\cite{Wood:1997zq} of atomic parity violation (APV), 
most notably in $^{133}$Cs where one also obtained~\cite{Porsev:2009pr,Dzuba:2012kx} 
the best understanding of atomic structure~\cite{Ginges:2003qt} 
which is crucial for the interpretation in terms of EW physics. 
The result,
\be\label{cs}
188\, g_{AV}^{\, e u} + 211\, g_{AV}^{\, e d} = 36.35 \pm 0.21\ ,
\ee
is $1.5~\sigma$ lower than the SM prediction of 36.66.
In the future, one may constrain different linear combinations by studying neutron rich nuclei like
Fr or Ra, or by considering isotope ratios in which atomic physics uncertainties cancel. 

Note that we are using the notation introduced in Reference~\cite{Erler:2013xha}.
More familiar are the so-called weak charges, $Q_W$, which at the tree level
are also given by the coherent sum of the corresponding quark couplings (as in Equation~\ref{cs}),
but multiplied by a factor of $-2$ and with a different set of radiative corrections applied.
This is indicated by the use of $C_{1q}$ and $C_{2q}$ in place of $g_{AV}^{\, e q}$ and $g_{VA}^{\, e q}$.
Our notation serves as a reminder that corrections have been applied (see Section~\ref{EFT}) 
allowing direct comparison and combination of the couplings when extracted from different
observables, experiments, and kinematic conditions.

PVES offers an alternative to APV and the possibility to cross-examine Equation~\ref{cs}
with entirely different experimental and theoretical challenges and uncertainties.
E.g., the Qweak Collaboration~\cite{Androic:2013rhu} has measured the left-right 
asymmetry~\cite{Cahn:1977uu} in elastic polarized $ep$ scattering, ${\vec{e}}^{\, -} p \rightarrow e^- p$,
\be
A_{LR}^{ep} \equiv
\frac{d\sigma_L - d\sigma_R}{d\sigma_L + d\sigma_R} =
- {s \over v^2} {g_{AV}^{\, ep} \over 4 \pi \alpha}  \left[ y + {\cal O}(y^2) \right] {\cal F}^{ep}_{\rm QED} (Q^2,y)\ ,
\ee
where $s = Q^2/y$ is the center-of-mass energy, $\alpha$ the EM fine structure constant, 
and ${\cal F}^{ep}_{\rm QED} (Q^2,y)$ is a QED correction factor.
Here and in the following, $y$ is the fractional energy transfer from the electrons to the hadrons
which is perturbatively small for Qweak kinematics, $y \approx 0.0082$, and will be smaller yet for 
the experiment at the MESA facility in Mainz.
The analysis of the Qweak commissioning data (about 4\% of the total) translates into the constraint,
\be
\label{QweakI}
2\, g_{AV}^{\, e u} + g_{AV}^{\, e d}  = - 0.032 \pm 0.006\ ,
\ee
to be compared to SM prediction of $-0.0355$.
The final result is expected to be four times as precise, and even greater precision will be possible in Mainz.
If additionally elastic scattering off isoscalar nuclei like $^{12}$C was used to extract $g_{AV}^{\, ep} + g_{AV}^{\, en}$,
one could disentangle $g_{AV}^{\, eu}$ and $g_{AV}^{\, ed}$ from PVES alone.

The analogous asymmetry in M\o ller scattering~\cite{Derman:1979zc},
\be\label{ALRee}
A_{LR}^{ee} = 
{s \over v^2} {g_{VA}^{\, ee} \over 4 \pi \alpha} {2 y (1 - y) \over (1 - y + y^2)^2}\ {\cal F}^{ee}_{\rm QED} (Q^2,y)\ ,
\ee
has been obtained  by the SLAC--E158 Collaboration~\cite{Anthony:2005pm} from which one can extract~\cite{Erler:2013xha},
\be
\label{SLAC-E158}
g_{AV}^{ee} = 0.0190 \pm 0.0027\ ,
\ee
while the SM predicts the value 0.0225.
The MOLLER Collaboration~\cite{Mammei:2012ph} at Jefferson Lab
aims to reduce the uncertainty in $g_{AV}^{ee}$ by a factor of five by taking advantage of the energy upgraded CEBAF.

Deep inelastic PVES (PVDIS) experiments~\cite{Prescott:1979dh,Zheng:2013} are sensitive 
to the interference of the quark-level amplitudes corresponding to ${\cal L}_{\rm NC}^{\, e f}$ with photon exchange.
Scattering from an isoscalar target provides information on the charge weighted combinations, 
$2 g_{AV}^{\, eu} - g_{AV}^{\, ed}$ and $2 g_{VA}^{\, eu} - g_{VA}^{\, ed}$.
In the quark model and in the limit of zero nucleon mass one can write~\cite{Erler:2013xha} 
in a simple valence quark approximation,
\be\label{eDIS}
A_{LR}^{{\rm DIS}} = - {3 \over 20 \pi \alpha(Q)} {Q^2 \over v^2} \left[ ( 2 g_{AV}^{eu} - g_{AV}^{ed} ) + 
( 2 g_{VA}^{eu} - g_{VA}^{ed} ) {1 - (1 - y)^2 \over 1 + (1 - y)^2} \right].
\ee
One has to correct for higher twist effects, CSV,
sea quarks, target mass effects, 
longitudinal structure functions and nuclear effects.
Some of these issues are of considerable interest in their own right, 
and their discussion is deferred to Sections~\ref{sechadronic} and~\ref{secnuclear}.

The proposed Electron Ion Collider (EIC)~\cite{Accardi:2012qut},
currently under study as the next step in exploring the QCD frontier,
is a high luminosity ($\sim 10^{33-34}\>$cm$^2$ s$^{-1}$) machine that
will use highly polarized ($\sim 70\%$) electron and nucleon beams
with a variable center of mass energy $20~{\rm GeV} < \sqrt{s} < 150~{\rm GeV}$ 
and a wide variety of nuclear targets. Such a facility
will allow further improved precision studies of PVDIS asymmetries.
Its wide kinematic range, along with other experiments, will allow one to disentangle  various
hadronic effects such as CSV or higher twist effects. 
In turn, it can also allow for more precise extractions of
the contact interaction couplings that are sensitive to new physics
and the weak mixing angle at different values of $Q^2$.

We will return to the EW couplings, their radiative corrections and the associated uncertainties
in Section~\ref{sectheory}.

\subsection{Hadronic and Nuclear Physics}

As discussed earlier, sound theoretical control over various hadronic effects that contribute to the asymmetries 
is essential for a reliable interpretation of the EW physics. 
Alternatively, one can view PVDIS as a probe of the hadronic effects themselves as a means 
to further our understanding of QCD and nuclear dynamics. 
Within this context, we discuss some of the important hadronic effects 
that affect PVDIS asymmetries in Section~\ref{sechadronic}.

We describe elastic PVES from heavy nuclei in Section~\ref{secnuclear}.  
These experiments can precisely locate neutrons in the nucleus 
because the weak charge of a neutron is much larger than that of a proton.  
These experiments target neutron radii that have important implications 
for nuclear structure, astrophysics and APV.  

\section{EXPERIMENTAL ISSUES}


\subsection{Introduction}

A large number of PVES experiments have been completed or are in progress.  
A list is given in {\bf Table~\ref{tab:expts:E}} which gives a flavor 
of the variety of both apparatus and physics goals.    
We will first give a brief description of selected experiments.  
Then we will discuss general design criteria that apply to all of the experiments.

\begin{table}[!t]
\caption{Apparates for selected PV experiments.  For magnets, Q (D) refers to quadrupole (dipole).}
\begin{center}
\begin{tabular}{c|c|c|c|c|c} \hline
Experiment & Magnets & Detector & Count & Angles & Physics \\ \hline
SLAC--E122 &Dipoles&Pb Glass&No&12$^\circ$& $g_{AV}^{eq}$, $g_{VA}^{eq}$ \\ \hline
Mainz &None & Air C& No&146$^\circ$&$g_{AV}^{eq}$ \\ \hline
MIT--Bates & Q & Lucite C &No & 20$^\circ$& $g_{AV}^{eq}$\\ \hline
SAMPLE &None&Air C&No &146$^\circ$ & Strange FF \\ \hline
HAPPEX & QQDDQ& Pb-Lucite & No &15$^\circ$ & Strange FF\\ \hline
G\O\ &Toroid&Scintillator&Yes  &20$^\circ$-50$^\circ$&Strange FF \\ \hline
Mainz--A4 &None&PbF$_2$&Yes &30$^\circ$&Strange FF \\ \hline
SLAC--E158 &QQQQ& Cu-Quartz & No & 2$^\circ$ & $g_{AV}^{ee}$ \\ \hline
HAPPEX--He &QQDDQ& Cu-Quartz & No & 5$^\circ$ & Strange FF \\ \hline
PREX &QQDDQ& Quartz & No & 5$^\circ$ & $R_n$\\ \hline
PREX-II &QQDDQ& Quartz & No & 5$^\circ$ & $R_n$\\ \hline
CREX  &QQDDQ& Quartz & No & 4$^\circ$ & $R_n$\\ \hline
JLab--Hall A &QQDDQ&Pb Glass&Yes &19$^\circ$& $g_{VA}^{eq}$ \\ \hline
Qweak & Toroid & Pb-Quartz & No & 5$^\circ$ & $g_{AV}^{eq}$\\ \hline
MOLLER &Toroid&Quartz&No&1$^\circ$&$g_{AV}^{ee}$ \\ \hline
SoLID &Solenoid&Package &Yes &22$^\circ$-35$^\circ$& $g_{VA}^{eq}$, CSV, HT \\ \hline
Mainz--P2 &Solenoid&Quartz &No&20$^\circ$&$g_{AV}^{eq}$\\ \hline
Mainz--C &Solenoid&Quartz &No&40$^\circ$&$g_{AV}^{eq}$ \\ \hline
\end{tabular}
\label{tab:expts:E} 
\end{center}
\end{table}

\subsection{Brief Descriptions of Selected Experiments}

In the Qweak experiment~\cite{Androic:2013rhu} electrons scattered from a 35~cm long LH$_2$ target 
with angles between 5$^\circ$ and 11$^\circ$ and passed through a collimator.  
They were deflected by a toroid and focussed onto quartz bars that served as Cherenkov detectors.  
Virtually all of the apparatus for Qweak was custom fabricated.
 
By contrast, the PVDIS~\cite{Zheng:2013} and PREX~\cite{Abrahamyan:2012gp} experiments both used the same apparatus, 
the JLab--Hall~A high-resolution spectrometers (HRS)~\cite{Alcorn:2004sb}. 
The HRS was designed not for PVES but rather to measure cross sections with excellent energy resolution.  
For that purpose, the HRS spectrometers were instrumented with a detector package with drift chambers for tracking, 
Cherenkov counters and an electron calorimeter made of Pb glass for particle identification.

For PREX, the spectrometers needed sufficient energy resolution to reject inelastically 
scattered electrons that had lost more than a few MeV.  
The HRS detector package was not used, but rather elastic events were focussed 
onto a small quartz bar and the signal was integrated.  
The drift chambers in the spectrometers were used in special calibration runs 
to measure the average $Q^2$ of the events giving signals in the quartz.  

For PVDIS, by contrast, energy resolution was irrelevant.  
However, Cherenkov and Pb Glass detectors were used to identify the DIS electrons, 
separating them from the more copious pions.  
Since the electrons were identified by a coincidence between two detectors, the events must be counted.

Many PVES experiments are built around magnets originally designed for a different purpose.
E122~\cite{Prescott:1978tm}, MOLLER~\cite{Mammei:2012ph}, 
Bates~\cite{Souder:1990ia}, SoLID~\cite{Souder:2008zz}, and P2~\cite{Becker:2013fya} are examples.  
Another variation are experiments like A4~\cite{Baunack:2009gy} 
and SAMPLE~\cite{Spayde:1999qg}, where there is no magnet but special detectors 
are designed that can detect the elastic events amidst a large background of lower energy particles.

\subsubsection{Spectrometers:}
The optimization of PVES experiments is as follows.
The differential cross sections and PV asymmetries are approximately, 
\be
\frac{d\sigma}{d\Omega}\approx\frac{E^2[FF]^2}{Q^4}\ , \qquad\qquad
A_{LR} = A_0 Q^2\ ,
\ee
where $A_0$ is typically 10$^{-4}$ to $10^{-5}$ GeV$^{-2}$, depending on the reaction.  
At first glance, it may appear that higher energies are better.  
However, for small angles, where $\theta \approx \sin\theta$,
we can show that the statistics are independent of $E$.
The reason is that since $A_{LR}$ depends strongly on $Q^2$,
the angular acceptance $\Delta\theta$ cannot be too large; 
typically $\Delta\theta\sim 0.2 \times\theta$.
At fixed $Q^2$, $\theta\sim 1/E$.  
Then, the solid angle  
$\Delta\Omega = \sin\theta \Delta\theta \Delta\phi \approx \theta \Delta\theta \Delta\phi$,
and $E^2\Delta\Omega$ is approximately constant.  
The statistical error varies as the ratio $\sigma/A_{LR}^2$, which is independent of $Q^2$.  
Thus, the main criteria for designing an experiment is to maximize the acceptance $\Delta\phi$.

Toroids, such as the one used for Qweak,  typically have a large $\Delta\phi \approx \pi$.
For MOLLER, where there are two electrons in the final state, full $\phi$ coverage can be obtained with a toroid.  
With a solenoid, the full $2\pi$ acceptance in $\phi$ can be achieved~\cite{Souder:1990ie}, 
as is planned for the P2 experiment.  
The JLab HRS spectrometers have a fixed solid angle acceptance.  
As a consequence, they are used at the most forward angles possible where they 
achieve a respectable $\Delta\phi$ of about $\pi/2$.

Another important requirement on the spectrometer, especially for experiments with elastic scattering, 
is energy resolution, which must be good enough to reject inelastic events.  
For proton scattering, pion production is the first background, so the resolution must be better than 100 MeV.
For nuclear targets, inelastic levels at the 5 MeV levels must be rejected.  
Often a low beam energy is optimal because the rate is independent of energy 
but the absolute resolution improves with lower energies.  
By contrast, for the JLab HRS spectrometers with their excellent energy resolution, backgrounds can be rejected 
even with high energies and rates go up with energy due to the fixed solid angles of the HRS system.

\subsubsection{Detectors:}
One important decision that influences both the design of the spectrometer 
and the detectors is whether to count events or to integrate the signal.
Integration has the advantage that there are no pile-up or dead-time corrections.  
E.g., with the high resolution spectrometers in Hall~A at JLab, elastically scattered events are physically separated 
from all inelastic events, and very clean data samples are obtained even though the signal is integrated.

For a counting experiment, the statistical error in $A_{LR}$ is simply $1/\sqrt{N}$, 
where $N$ is the number of events detected.  
For an integrating experiment with average signal $S$ and detector resolution $\sigma$, 
the statistical error is given by,
\be
\delta A_{LR} = \sqrt{\frac{1+\sigma^2/S^2}{N}}\ .
\ee
Thus, if the detector has reasonable resolution, little statistics are lost by integrating.  
For example, for a moderate resolution $\sigma/S \sim 20\%$, the statistical error is increased by only 2\%.
However, care must be taken.  
Half of the statistics would be lost if 1\% of the events has 10 times the average signal.  
With a thin scintillator, the Landau tail would  causes $\sigma/S$ to be large, 
and integrating the signal would cause unacceptable loss in statistics. 
For high energy electrons, a crude shower counter with coarse granularity will suffice.
For high rates, radiation damage becomes important, and fused silica (henceforth called quartz) is the material of choice.
For energies of 1~GeV and below, the resolution of a granular detector becomes poor, 
and thin quartz has better performance.  
The design is critical.  
If the quartz is too thin, there are too few photoelectrons detected and $\sigma$ increases.  
If the quartz is too thick, the electrons start to shower and create a high-signal tail, which increases $\sigma$.  
With great care in optimizing the collection of light, acceptable performance can be achieved. 

For experiments with larger asymmetries and thus lower event rates, counting techniques are practical, 
taking advantage of the continuos wave nature of modern electron facilities.  
For SoLID, a traditional electron spectrometer with tracking, Cherenkov counter, and a calorimeter is proposed. 

\subsubsection{Data Acquisition and Electronics:}
Perhaps the most specialized aspect of PVES is the electronics.   
Since integration is only used for PVES, the integrating electronics is custom made.  
For the high rate experiments, such as SLAC--E158, PREX, and Qweak, 
the problem is that the statistical noise in a given helicity window is only on the order of 100 ppm, 
and keeping the noise of the electronics below that level is a special challenge.

For the counting experiments, the large rate of accepted events separates PVES experiments from other experiments.  
Custom counting electronics were a main feature of the A4 program, the JLab PVDIS experiment, and SoLID.
  
\subsubsection{Backgrounds:}
Backgrounds are an  important source of error.  
The easiest case is elastic scattering where the highest energy particles possible are the desired events, 
and backgrounds can be eliminated simply by achieving good resolution, as is done with the JLab--Hall~A HRS spectrometers.  
For DIS, pions are the main background, and can be separated by the usual methods.

General electromagnetic background can be a problem, especially for toroidal spectrometers.
Helicity-dependence of the beam width, or other higher-order parameters which are very difficult to monitor,
can propagate to the background if it arises from beam spraying from a small collimator.
There can be physics asymmetries in some backgrounds, such as decay products from hyperons or pions produced by DIS events.  
Detailed studies are required to untangle these effects

\subsection{Precision of PVES Experiments}

As the field of PVES has developed, the precision achieved or proposed, 
both in terms of the absolute size of the error and of the fractional error in the asymmetry,
has improved dramatically, as shown in {\bf Table~\ref{tab:expts:A}}.  
Most of the techniques that have been perfected over the years are common to all PVES experiments.

\begin{table}[!t]
\caption{Results from selected PV experiments. 
Asymmetries are given in ppm.}
\begin{center}
\begin{tabular}{c|c|c|c|c|c|c} \hline
Experiment & Reference &Year & $-A_{LR}$ & $\delta A$ (stat) & $\delta A$ (syst) & $\delta A$/$A$(\%)\\ \hline
SLAC--E122 &\cite{Prescott:1979dh}&1978&$-120$&7&6&8\\ \hline
Mainz &\cite{Heil:1989dz}&1989& $-9.4$& 1.8& 0.5&20 \\ \hline
MIT--Bates &\cite{Souder:1990ia}& 1990 & 1.62 & 0.37 & 0.11 &24\\ \hline
SAMPLE &\cite{Spayde:1999qg}&1990 & $-5.61$&0.67&0.88&20 \\ \hline
HAPPEX & \cite{Aniol:2004hp}& 2001 & $-15.05$ & 0.98 & 0.56& 7.5\\ \hline
G\O\ &\cite{Armstrong:2005hs}&2005&$-2$&0.15&0.2&13 \\ \hline
Mainz--A4 &\cite{Baunack:2009gy}&2009&$-17.2$&0.8&0.9&5 \\ \hline
SLAC--E158 &\cite{Anthony:2005pm}& 2005 & $-0.131$ & 0.014 & 0.010&13 \\ \hline
HAPPEX--He &\cite{Acha:2006my}& 2007 & 6.40 & 0.23 & 0.12 &4.1\\ \hline
PREX &\cite{Abrahamyan:2012gp}& 2012 & 0.656 & 0.060 & 0.014 &9.4\\ \hline
PREX-II&& 2016 & (0.51) & 0.015 & 0.005 &3\\ \hline
CREX && 2017 & 2.0 & 0.04 & 0.02 &2.4\\ \hline
JLab--Hall A &\cite{Zheng:2013}&2014&$-160$&6.4&3.1&4.4 \\ \hline
Qweak & \cite{Androic:2013rhu} & 2013 &$-0.280$ & 0.035 & 0.031& 17\\ \hline
MOLLER &&2020&0.035&0.0007&0.0004&2.3 \\ \hline
SoLID &\cite{Souder:2008zz}&2022&$-800$&2&4&0.6 \\ \hline
Mainz--P2 &\cite{Becker:2013fya}&2018 &$-0.020$&0.00025&0.00019&1.7 \\ \hline
Mainz--C &&2020 &0.8&0.0025&0.00017&0.3 \\ \hline
\end{tabular}
\label{tab:expts:A}
\end{center}
\end{table}

\subsubsection{Measuring Small Asymmetries:}
The basic technique used to observe the small asymmetries in PVES
is the ability to rapidly reverse the helicity of the beam without changing any of its other properties, 
including energy, intensity, position, and angle.  
Many systematic errors that are important for measurements of cross sections, such as target thickness, 
spectrometer solid angle, etc., which are hard to measure and tend to drift with time, cancel in the asymmetry.

The source of the polarized electrons in PVES is photo-emission from a crystal, 
and the helicity of the beam is reversed by reversing that of the laser light producing the photo-electrons.  
Today, polarizations of more that 85\% and beam currents of more than 100 $\mu$A are now routinely achieved 
by using a strained GaAsP crystal~\cite{Sinclair:2007ez,Adderley:2010zz}.

The helicity of the electron beam is determined by that of the laser light producing the photo-emission.  
The helicity of the light is reversed by using a device called a Pockels cell, 
which is a crystal whose birefringence is controlled by high voltage applied across the cell.  
To first order, the helicity is the only beam property changed by the voltage on the Pockels cell. 

The improvement in the achievable precision in PVES  is in part due to great progress 
in understanding and correcting the imperfections in the helicity reversals.  
The basic problem is that the laser light is partially linearly polarized, and the linear polarization also reverses with helicity.  
Flipping the linear component of the light causes systematic differences in the intensity, position and width 
of the photo-emitted electrons, the latter effects arising from a spatial dependence of the linear polarization.  
Extensive studies of these effects resulted in specialized techniques 
that have greatly improved the ability to provide clean helicity reversals.

Systematic errors can further be reduced by using independent means to reverse the beam helicity.
These include insertion of a half wave plate in the laser beam, 
using a Wien filter in the electron beam before it is accelerated, 
and sometimes by running the experiment at a nearby energy 
where the helicity is reversed by an extra $g-2$ flip in the accelerator or beam transport.  
The half-wave plate reversals can be done every few hours; 
the other reversal methods are done less frequently because the accelerator needs to be retuned after the reversal.

The spectrometers in PVES experiments are designed so that they are as insensitive as possible to the beam parameters.   
By making the apparatus symmetrical, the signal is very insensitive to differences in beam position and angle.  
Careful design of apertures can reduce the sensitivity to beam size.  
In addition, careful tuning of the accelerator reduces the size of the helicity-correlated
beam differences at the target. 

Systematic differences in the beam parameters are controlled by a set of position-sensitive
monitors that together determine the beam position and angle on target.  
A position monitor in a point of high dispersion determines energy differences in the beam. 
The sensitivity of the spectrometer to these beam parameters can be measured by dithering the beam 
with coils and an RF cavity so corrections can be made for any residual beam differences~\cite{Kumar:2000eq}.  
Feedback is another tool, especially useful for eliminating the helicity-dependence of the beam intensity.

\subsubsection{Scale errors:}
Although the asymmetries measured in PVES experiments are small, in many cases the fractional error  
$\delta A_{LR}/A_{LR}$ is on the order of 5\% and the SoLID experiment proposes to measure 
$\delta A_{LR}/A_{LR}$ to 0.5\%, which is a higher precision than most cross section experiments. 
Systematic errors in scale factors, such as the beam polarization $P_e$ and $Q^2$, become important.

There are two methods for measuring $P_e$, each with variations.  
One is Compton scattering from a polarized laser beam.  
Either the scattered electron or photon can be detected; 
indeed both can be detected to provide independent measurements.  
Theoretical errors for Compton scattering are small, 
and the beam  polarization can be measured simultaneously with $A_{LR}$.  
Precisions as good as 1\% have been achieved and an additional factor of two is possible.

The other method is M{\o}ller scattering from a target with polarized electrons.  
For a ferromagnetic material in a low magnetic field, errors in the target polarization are on the order of 3\%.  
By saturating the target in a 4~T field, this error can be reduced to below 1\%.  
Unfortunately, the targets are thick and can tolerate currents on the order of a few $\mu$A 
and the polarimetry cannot be done while $A_{LR}$ is measured.  
A proposed variation~\cite{Chudakov:2004de}, which uses cold H atoms in a magnetic trap, 
features a thin target that can be run during the $A_{LR}$ measurement, 
and promises to provide precision below 0.5\%

\section{THEORETICAL ISSUES}
\label{sectheory}


\subsection{Effective Couplings}
\label{EFT}
In an effective field theory one absorbs the effects 
of the heavy degrees of freedom, such as the $W$, the $Z$, and the Higgs boson
mediating the weak interaction, into effective couplings.
The degrees of freedom of the effective theory relevant to PVES are then electrons,
first generation quarks, and nucleons.  
At the level of radiative corrections, discussed in Section~\ref{radcorr},
it is important to define the couplings in a process-independent manner and for an arbitrary gauge theory, 
so as to allow for direct comparisons and global analyses of the different types of experimental information.
We use the conventions proposed in Reference~\cite{Erler:2013xha} which aim to strike a balance between
formal, practical and historical considerations.

After EW gauge symmetry breaking, the SM fermion fields $\psi_f$ (with mass $m_f$) interact 
with the Higgs field $H$, the intermediate vector bosons $W^\pm$ and $Z$, 
and the photon field $A$, according to the Lagrangian
\be\label{LfY}
{\cal L} = - \sum_f {m_f \over v} \ovl \psi_f \psi_f H - {g \over \sqrt{2}} 
\left[ {J_W^\mu}^\dagger W_\mu^+ + J_W^\mu W_\mu^- + J_A^\mu A_\mu + J_Z^\mu Z_\mu \right]\ ,
\ee
where $v$ is the vacuum expectation value of $H$.
We have chosen a common normalization in which
the charged (CC), electromagnetic (EM) and weak neutral currents (NC) are given by,
\be
J_W^\mu = \ovl{d_L} \gamma^\mu V_{\rm CKM}^\dagger u_L + \ovl{e_L} \gamma^\mu \nu_L\ , 
\ee
\be
J_A^\mu = \sqrt{2}\,  \sin\theta_W (Q_u \ovl{u} \gamma^\mu u + Q_d \ovl{d} \gamma^\mu d + Q_e \ovl{e} \gamma^\mu e)\ , 
\ee
\be
J_Z^\mu = 
\sum_f \ovl\psi_f \gamma^\mu \left[ g_L^f P_L + g_R^f P_R \right] \psi_f =
\sum_f \ovl\psi_f \gamma^\mu {g_V^f - g_A^f \gamma^5 \over 2} \psi_f =
\ee
\be
{1 \over \sqrt{2}\, \cos\theta_W} (\ovl{u_L} \gamma^\mu u_L - \ovl{d_L} \gamma^\mu d_L + \ovl{\nu_L} \gamma^\mu \nu_L -
\ovl{e_L} \gamma^\mu e_L) - \tan\theta_W J_A^\mu\ .
\ee
Here, $\psi_{L,R} \equiv P_{L,R} \psi$ denote chiral projections, 
$J_W^\mu$ contains the quark mixing matrix $V_{\rm CKM}$, $\theta_W$ is the weak mixing angle,
and fermion generation indices have been ignored.
Furthermore,
\be\label{vzaz}
g_V^f \equiv g_L^f + g_R^f = \sqrt{2} {T^3_f - 2 \sin^2\theta_W Q_f \over \cos\theta_W}\ , 
\qquad\qquad
g_A^f \equiv g_L^f - g_R^f = \sqrt{2} {T^3_f \over \cos\theta_W}\ ,
\ee
are vector and axial-vector $Z$ couplings with $T^3_u = T^3_\nu = - T^3_d = - T^3_e = 1/2$. 

At low energies, $Q^2 \equiv -q^2 \ll M_{W,Z}^2$, one finds the effective four-fermion interaction Lagrangians,
\be\label{Leff}
{\cal L}_{\rm CC} = - {2 \over v^2} J^{\mu\dagger}_W J_{W\mu}\ , 
\qquad\qquad
{\cal L}_{\rm NC} = - {\cos^2\theta_W \over v^2} {J^\mu_Z} J_{Z\mu}\ .
\ee

\subsection{Radiative Corrections}
\label{radcorr}
The low-energy couplings defined in Equation~\ref{LeqNC} are modified by radiative corrections, 
which in general depend on energies, experimental cuts, etc. 
To render them universal, adjustments have to be applied to the underlying processes. 
One {\em defines\/}~\cite{Erler:2013xha} the one-loop radiatively corrected couplings to include the purely EW diagrams 
and certain photonic loops and $\gamma$-exchange graphs, while the remaining corrections,
as e.g., the interference of two photon exchange diagrams with single $\gamma$ or $Z$ exchanges,
are assumed to be applied individually for each experiment. 
Moreover, since the genuine EW radiative corrections will in general depend on the specific kinematical points or ranges 
at which the low-energy experiments are performed, one needs to introduce idealized EW coupling parameters~\cite{Erler:2013xha} 
defined at the common reference scale $\mu = 0$, and expect the experimental collaborations to adjust to their conditions.
According to the calculations from References~\cite{Marciano:1982mm,Marciano:1983ss,Czarnecki:1995fw,Erler:2003yk},
the SM expressions for the NC couplings defined in this way~\cite{Erler:2013xha} are given by,
\be\label{gAVef}
g_{AV}^{\, \ell f} = \rho \left[ - T^3_f + 2\, Q_f \hat s_0^2 -  2\, Q_f \diameter_{\ell Z} + 
\Box_{ZZ} + \Box_{\gamma Z} \right]  - 2\, Q_f \diameter_{\ell W} + \Box_{WW}\ ,
\ee
\be\label{gVAef} 
g_{VA}^{\, \ell f} = \rho \left[ - T^3_f (1 - 4 \hat s_0^2) + 2 \diameter_{fZ} + \Box_{ZZ}^\prime + 
\Box_{\gamma Z}^\prime \right] + 2\, \diameter_{fW} + \Box_{WW}\ ,
\ee
where using the abbreviations $\hat\alpha_{V} \equiv \hat\alpha(M_V)$ and 
$\hat\alpha_{ij} \equiv \hat\alpha(\sqrt{m_i M_j})$, one has,
\be
\diameter_{fW} = {\alpha \over 6 \pi} \left[ (Q_f - 2 T^3_f) \left( \ln {M_W^2 \over m_p^2} + {1\over 6} \right) - {8 \over 3} T^3_f \right]\ ,
\ee
\be
\diameter_{fZ} = {\alpha \over 6 \pi} Q_f\, g_{VA}^{\, ff} \left( \ln {M_Z^2 \over m_f^2} + {1\over 6} \right)\ ,
\vspace{6pt}
\ee
\be\label{WWbox}
\Box_{WW} = - {3 \hat\alpha_Z \over 16 \pi \hat s_Z^2} 
\left[ 1 - {\hat\alpha_s(M_W) \over \pi} + {2 \over 3}\, T^3_f \left( 5 - {\hat\alpha_s(M_W) \over \pi} \right) \right]\ ,
\vspace{6pt}
\ee
\be\label{ZZbox}
\Box_{ZZ} = - {3 \hat\alpha_Z \over 16 \pi \hat s_Z^2 \hat c_Z^2} 
\left( g_{VA}^{\, \ell f} g_{VV}^{\, \ell f} + g_{AV}^{\, \ell f} g_{AA}^{\, \ell f} \right)
\left[ 1 - {\hat\alpha_s(M_Z) \over\pi} \right]\ ,
\ee
(the QCD correction needs to be dropped if $f$ refers to a lepton) and $\Box_{ZZ}^\prime$ is given by $\Box_{ZZ}$ with $g_{VA}^{\, \ell f} \leftrightarrow g_{AV}^{\, \ell f}$.
Furthermore, 
\be\label{gammaZ}
\Box_{\gamma Z} = {3 \hat\alpha_{fZ} \over 2 \pi} Q_f\, g_{VA}^{\, \ell f} \left[ \ln {M_Z^2 \over m_f^2} + {3 \over 2} \right]\ ,
\hspace{13pt}
\Box_{\gamma Z}^\prime = {3 \hat\alpha_{pZ} \over 2 \pi} Q_f\, g_{AV}^{\, \ell f} \left[ \ln {M_Z^2 \over m_p^2} + {5 \over 6} \right]\ .
\ee
In these relations, the parameter $\rho$~\cite{Veltman:1977kh} renormalizes 
the NC interaction strength at low energies, and $\Box_{WW}$ and $\Box_{ZZ}$ denote EW box diagrams.
The logarithms entering the quark charge radii, $\diameter_{q W}$ and $\diameter_{q Z}$, 
are regulated at the strong interaction scale, $\Lambda_{\rm QCD}$, 
introducing a hadronic theory uncertainty into the $g_{VA}^{\, \ell q}$ unless they are extracted from DIS
(in these expressions $m_q = m_p$ by {\em definition\/}).

Numerically most important are vacuum polarization diagrams of $\gamma$-$Z$ mixing type 
giving rise to a scale-dependence~\cite{Czarnecki:2000ic} of the weak mixing angle.
As long as one stays in the perturbative QCD domain these effects can be re-summed~\cite{Erler:2004in}, 
while passing $\Lambda_{\rm QCD}$ introduces a hadronic uncertainty (see Section~\ref{uncertainties}).
At the tree level we employ the weak mixing angle at the scale $\mu = 0$ and abbreviate, 
$\hat s_0^2 \equiv \sin^2\hat\theta_W(0)$, while in the EW box graphs the use of
$\hat s_Z^2 \equiv \sin^2\hat\theta_W(M_Z)$ is more suitable
(the caret indicates the definition in the \msbar-renormalization scheme).
In $\diameter_{f Z}$ and in the $\gamma Z$ box diagrams discussed 
in Section~\ref{uncertainties} we use $\mu^2 = m_f M_Z$~\cite{Erler:2013xha}.  
The resulting SM values of the NC couplings are given in {\bf Table~\ref{ge}}.

\begin{table}[!t]
\caption{SM values of the one-loop and leading two-loop corrected 
effective NC couplings entering PVES for the charged SM fermions, the nucleons, and carbon,
where the latter cases refer to the coherent sums over constituent quark.}
\begin{center}
\begin{tabular}{c|c|c|c|c|c|c|c} 
\hline
$f$ & $e$ & $u$ & $d$ & $p$ & $n$ & $2u - d$ & $^{12}$C \\
\hline
$g_{AV}^{\ e f}$ &  0.0225 & $-0.1887$ &  0.3419 & $-0.0355$ &  0.4951 & $-0.7192$ & $\ph- 2.7573$ \\
$g_{VA}^{\ e f}$ &  0.0225 & $-0.0351$ &  0.0248 & $-0.0454$ &  0.0144 & $-0.0950$ & $      - 0.1859$ \\
\hline
\end{tabular}
\label{ge}
\end{center}
\end{table}

Unlike the quantities $C_{1q}$ and $C_{2q}$ defined~\cite{Marciano:1982mm} in the context of APV
we exclude the small axial current QED renormalization factors $(1 - Q_f^2 \alpha/2 \pi)$
from the definitions of $g_{AV}^{ef}$ and $g_{VA}^{ef}$, because 
QED and QCD corrections to external lines are not considered part of the EW couplings.
They can be included together with experiment specific initial and final state radiation effects and 
$\gamma\gamma$ box graphs in explicit QED radiative correction factors~\cite{Zykunov:2005md} 
multiplying the asymmetries.
In addition, $\diameter_{e Z}$ and the $\gamma Z$ box contributions need to be adjusted 
for $Q^2 \neq 0$, which in the case of M\o ller scattering amounts to shifts~\cite{Erler:2013xha}
\be\label{MOLLERcorr}
g_{VA}^{\, ee} \to g_{VA}^{\, ee} + 0.0010 \pm 0.0004\ , \qquad\qquad
g_{VA}^{\, ee} \to g_{VA}^{\, ee} + 0.0008 \pm 0.0005\ ,
\ee
for the SLAC~\cite{Anthony:2005pm} and Jefferson Lab~\cite{Mammei:2012ph} experiments, respectively.

In the DIS regime, the $\gamma Z$ box graphs need to be adjusted for the relevant $Q^2$-values 
and beam energies, but this is feasible~\cite{Arbuzov:1995id} and the extra $Q^2$-dependences 
are expected to change the $A_{LR}^{{\rm DIS}}$ by at most a few $\permil$.
Ignoring this issue, the adjustments for the SLAC~\cite{Prescott:1979dh} 
and Jefferson Lab~\cite{Zheng:2013,Souder:2012zz} experiments, 
all with $Q^2$-values around the charm quark threshold, are
\be
g_{AV}^{\, eq} \to g_{AV}^{\, eq}  - 0.0011\, Q_q \ln {Q^2 \over 0.14 \mbox{ GeV}^2}\ .
\ee
\be
g_{VA}^{\, eu} \to g_{VA}^{\, eu}  - 0.0009 \ln {Q^2 \over 0.078 \mbox{ GeV}^2}\ ,
\ee
\be
g_{VA}^{\, ed} \to g_{VA}^{\, ed}  + 0.0007 \ln {Q^2 \over 0.021 \mbox{ GeV}^2}\ .
\ee

\begin{figure}[!t]
\begin{center}
\vspace{-12pt}
\includegraphics[width=0.93\textwidth]{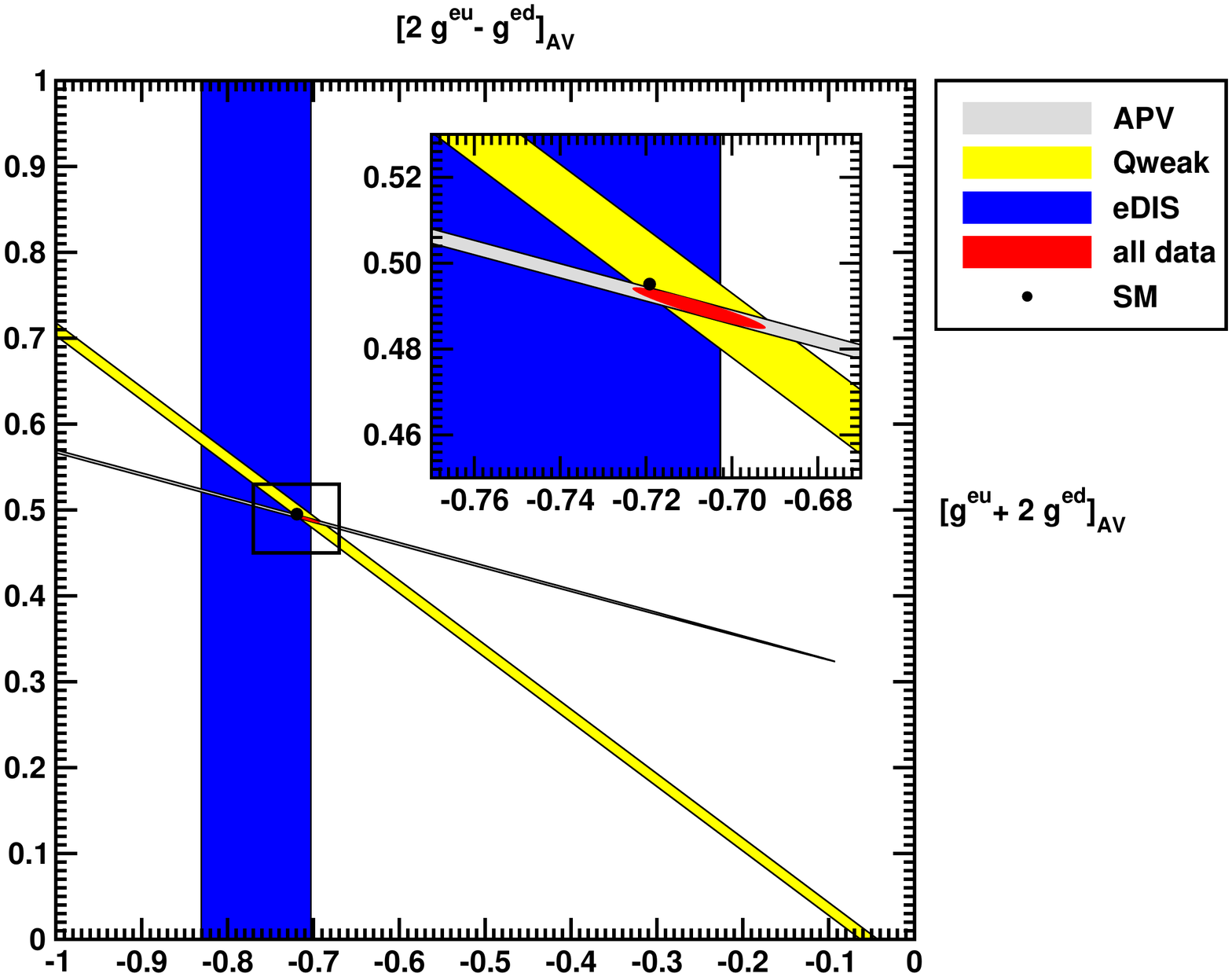}\vspace{-12pt}
\includegraphics[width=0.93\textwidth]{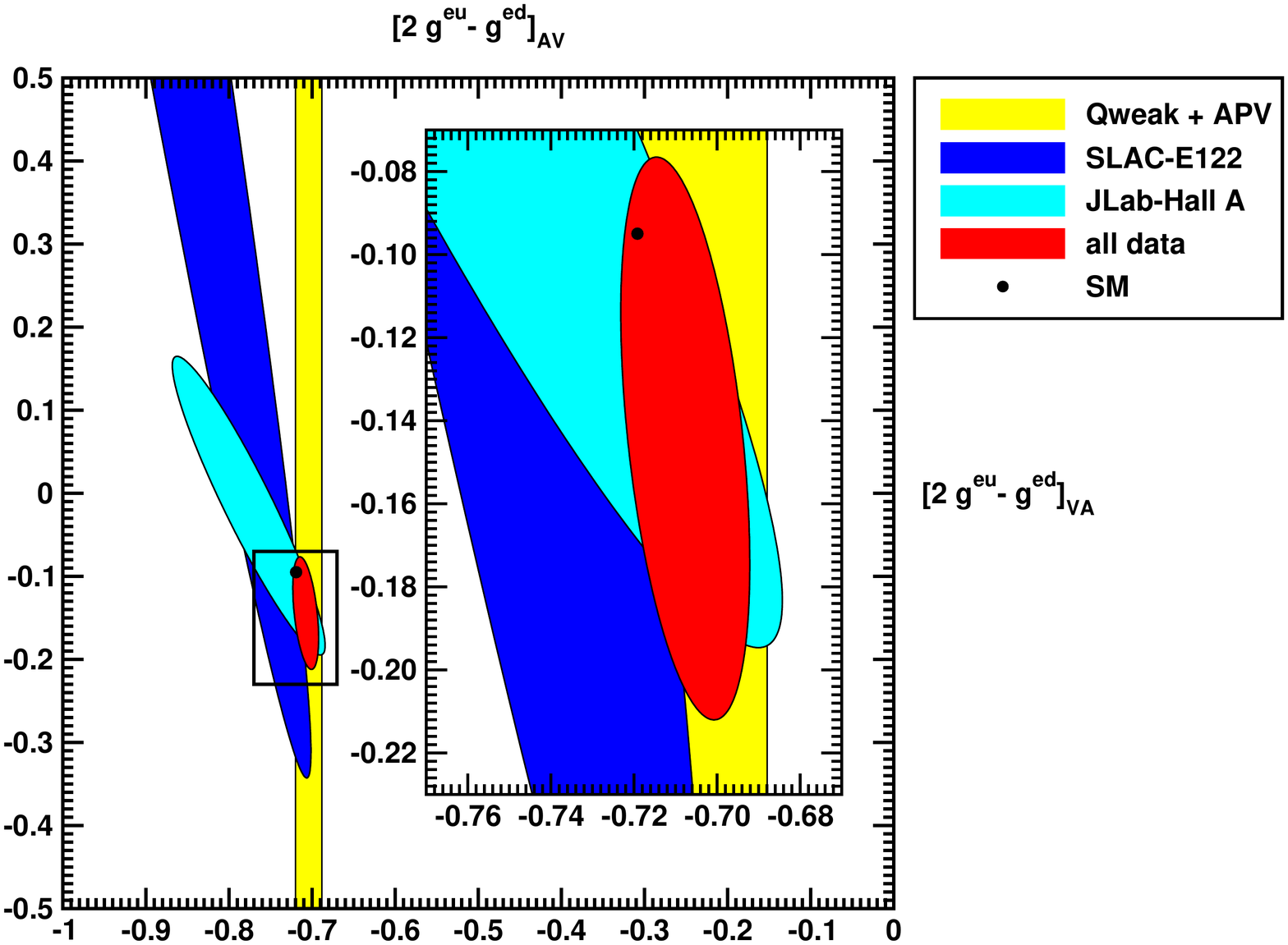}
\vspace{-12pt}
\caption{Experimentally determined coupling combinations $g_{AV}^{en}$ (upper plot) and $2\, g_{VA}^{eu} - g_{VA}^{ed}$ (lower plot)
vs. $2\, g_{AV}^{eu} - g_{AV}^{ed}$ compared to the SM prediction.
The APV constraint is from Cs and Tl and forms a strongly elongated ellipse rather than a band.
}
\vspace{-12pt}
\label{contours}
\end{center}
\end{figure}

The most precise current constraints on the corrected $g_{AV}^{en} \equiv g_{AV}^{eu} + 2\, g_{AV}^{ed}$ and 
$2\, g_{AV}^{eu} - g_{AV}^{ed}$ as functions of $2\, g_{VA}^{\, eu} - g_{VA}^{\, ed}$ are shown in {\bf Figure~\ref{contours}}.

\subsection{Theoretical Uncertainties}
\label{uncertainties}

The corrections that need to be applied in order to extract the coefficients 
$g_{AV}^{\, ef}$ and $g_{VA}^{\, ef}$ introduce additional uncertainties. 
These are relatively enhanced whenever the tree-level expression is suppressed
by a factor $1 - 4 \hat{s}_0^2 \approx 0.045$, as is the case for all the $g_{VA}^{\, ef}$,
as well as the $g_{AV}^{\, ef}$ for the electron and the proton. 
The most important associated theoretical uncertainties are from 
the renormalization group evolution (running) of the weak mixing 
angle~\cite{Czarnecki:2000ic,Erler:2004in} and
$\gamma Z$-box diagrams (both involving hadronic effects), 
as well as from unknown higher order EW corrections.

\subsubsection{$\sin^2 \hat\theta_W(0)$:}
The running of $\sin^2 \hat\theta_W$ from $\mu = M_Z$ 
(where it has been measured precisely in $Z$-pole experiments) to $\mu = 0$
arises from loop-induced $\gamma$-$Z$ mixing diagrams and amounts to a 3.2\% effect.
There is currently a relative 0.9\% theoretical uncertainty in its calculation~\cite{Erler:2004in} 
which translates to a 0.6\% uncertainty in the M\o ller asymmetry 
and a 0.4\% uncertainty for the $ep$ asymmetry
(it would be a negligible 0.03\% for measurements in carbon).

Most of the scale evolution ($\sim 75$\%) can be computed reliably within perturbation theory,
but the contribution below $m_c$ requires other considerations.
In this regime one can estimate the effect by relating it to that of the EM coupling, $\hat\alpha(\mu)$~\cite{Erler:1998sy},
which in turn is determined by computing a dispersion integral over $e^+ e^-$ cross-section
data and (up to isospin breaking effects) to $\tau$ decay spectral functions~\cite{Davier:2010nc}. 
The corresponding error contribution is about $\pm 3\times 10^{-5}$ in $\hat{s}_0^2$ 
and fully correlated with the one in $\hat\alpha$.

This strategy is limited by the necessary separation of the contributions from the various quark flavors 
and from QCD annihilation (or singlet) diagrams which enter $\hat\alpha$ and $\sin^2 \hat\theta_W$ differently.
The largest uncertainty ($5\times 10^{-5}$) is induced by the need to quantify the strange quark contribution
relative to the first generation quarks.  
In the future it may be possible to decrease this error by including information about the strange meson 
production fraction, but this is complicated due to the occurrence of secondary strange pair production.
Isospin breaking due to $m_u \neq m_d$ contributes an additional but much smaller ($\sim 10^{-5}$) uncertainty.
Currently an error of $3\times 10^{-5}$ is assigned to the singlet separation~\cite{Erler:2004in}, 
but since QCD annihilation diagrams are related to Okubo-Zweig-Iizuka (OZI) rule violations 
which are phenomenologically known to be strongly suppressed, this is quite likely an overestimate.

In addition, there are parametric uncertainties from the imperfect knowledge of $\alpha_s$, $m_b$ and $m_c$.
They could increase the above error estimates by up to 30\%,
but it can be expected that they will be much better determined in the future,
e.g., by means of lattice simulations.
In any case, these are fundamental fit parameters allowed to vary in EW fits, 
and should not be added to the purely theoretical uncertainties reviewed here.

\subsubsection{Hadronic $\gamma Z$ box:}
From Equation~\ref{gammaZ} one can see that the contributions from $\gamma Z$ box diagrams are logarithmically enhanced.  
Moreover, and more importantly, these logarithms are sometimes regulated at the scale $\Lambda_{\rm QCD}$
(we chose the reference value $m_p$ while the full effect depends on experimental and kinematical details).
This signals a hadronic uncertainty and is indicated 
when the parton model expression introduces the logarithm of the mass of a quark. 
Fortunately, the sum of uncrossed and crossed $\gamma Z$ box diagrams entering APV gives rise to 
the chiral structure $\Box_{\gamma Z}$ which is itself suppressed by $1 - 4 \hat{s}_0^2$.
This is the reason why this kind of uncertainty is small in APV~\cite{Marciano:1983ss,Blunden:2012ty}.

The finite beam energy, $E_e$, in PVES, on the other hand, upsets the cancellation
between uncrossed and crossed $\gamma Z$ box graphs, effectively introducing 
the wrong-chirality and unsuppressed structure $\Box'_{\gamma Z}$~\cite{Gorchtein:2008px}.
Several groups estimated the effect and there is consensus regarding the central value.  
The most recent evaluation~\cite{Rislow:2013vta} of the sum of the two chiral structures 
applicable to the Qweak condition with $E_e = 1.165$~GeV implies the correction,
\be\label{gammaZqweak}
g_{AV}^{\, ep} \to g_{AV}^{\, ep} - 0.0027 \pm 0.0007 \mbox{ (CEBAF)}\ .
\ee
The uncertainty is dominated by the $\Box'_{\gamma Z}$ structure and, respectively, smaller and larger 
by more than a factor of two compared to the ones quoted in~\cite{Gorchtein:2011mz} and~\cite{Hall:2013hta}.
A reduction and robust estimate of the error will be important 
for a solid interpretation of the Qweak experiment~\cite{Gorchtein:2013ila}.
Further hadronic effects of relative order $Q^2$ and their uncertainties ($\approx 1.5\%$) 
are treated by extrapolating the Qweak and other PVES data points to $Q^2 = 0$,
as e.g., in Reference~\cite{Young:2007zs}.
The $\gamma Z$ box corrections, uncertainties, and correlations 
need to be applied to each data point in the extrapolation.

The beam energy of MESA for the P2 project is still subject to optimizations, 
but it will be low enough that the $\Box_{\gamma Z}$ uncertainty dominates.
For $E_e = 200$~MeV, the analysis of Reference~\cite{Rislow:2013vta} implies the correction,
\be\label{gammaZP2}
g_{AV}^{\, ep} \to g_{AV}^{\, ep} - 0.0008 \pm 0.0003 \mbox{ (MESA)}\ .
\ee

Effects due to $Q^2 \neq 0$ in the shifts~\ref{gammaZqweak} and~\ref{gammaZP2}
are negligible~\cite{Gorchtein:2011mz}, and the $Q^2$-dependence of the weak mixing angle 
can be ignored if the asymmetry is normalized using the fine structure constant in the Thomson limit. 
But the electron charge radius induces the additional shift~\cite{Erler:2013xha},
\be\label{cradqweak}
g_{AV}^{\, ep} \to g_{AV}^{\, ep}  - 0.00008 \ln {Q^2 \over 0.00021 \mbox{ GeV}^2}\ .
\ee

In the DIS regime, the $\gamma Z$ box can be calculated perturbatively,
but the appropriate event generators~\cite{Kwiatkowski:1990es} need to be examined for consistency
with more recent conventions and refinements.
This will introduce an additional $Q^2$-extrapolation. 

\subsubsection{Unknown higher orders:}
Several EW one-loop corrections are large, most notably the $WW$ box contribution to the $ep$ asymmetry
which exceeds the anticipated experimental precision by an order of magnitude. 
Thus, the $\alpha_s$ terms~\cite{Erler:2003yk} in Equations~\ref{WWbox} and~\ref{ZZbox},
as well as other higher order effects need to be included or induce additional uncertainties.
For example, the uncertainties displayed in Equation~\ref{MOLLERcorr} arise from scale uncertainties
in one-loop terms.
Some reducible contributions can be determined by renormalization group and similar techniques, 
but a full two-loop calculation (ideally including enhanced three-loop effects) appears feasible and should be
vigorously pursued~\cite{Aleksejevs:2013fd}. 


\section{HADRONIC STRUCTURE}
\label{sechadronic}


The idealized form of the DIS asymmetry for a deuteron target in Equation~(\ref{eDIS}),
is modified by various hadronic corrections to 
\be
\label{apv2}
A_{LR}^{\rm DIS}  = - \frac{3}{20\pi \alpha} \frac{Q^2}{v^2}\, \left[ \tilde{a}_1 + \tilde{a}_2 \frac{1-(1-y)^2}{1+(1-y)^2} \right]\ ,
\ee
where the parameters $ \tilde{a}_j\ (j=1,2)$ have the form
\be
\label{apv3}
\tilde{a}_j = (2 g^{eu}_{j} -g^{ed}_{j}) \left[1 + R_j({\rm sea}) + R_j({\rm CSV})+R_j({\rm TM})+R_j({\rm HT}) \right]\ ,
\ee
and where $g^{eq}_{1}\equiv g^{eq}_{AV}$ and $g^{eq}_{2}\equiv g^{eq}_{VA}$.
The quantities $R_j({\rm sea}), R_j({\rm CSV}), R_j({\rm TM})$ and $R_j({\rm HT})$ 
denote corrections arising from sea-quarks, charge symmetry violation, 
target mass effects, and higher twist effects, respectively. 
Below we discuss some of these hadronic effects and their potential impact 
on the theoretical interpretation of precision PVDIS measurements.

Within the SM, it is often convenient to write Equation~\ref{apv2} in terms of 
the five EW structure functions $F_{1,2}^{\gamma}$ and $F_{1,2,3}^{\gamma Z}$ as
\be
\label{apv1}
A_{LR}^{\rm DIS} = \frac{1}{8\pi \alpha} \frac{Q^2}{v^2}\,
\left[ Y_1 \frac{F_1^{\gamma Z}}{F_1^\gamma} +  {Y_3\over 2} \frac{F_3^{\gamma Z}}{F_1^\gamma} \right]\ .
\ee
Note, that deviating from common conventions, we have absorbed $g_A^e$ and $g_V^e$,
in the first and second term, respectively, into the structure function.
The functions $Y_1$ and $Y_3$ functions have the form
\bea
&& Y_1 = \left[ \frac{1+R^{\gamma Z}}{1+R^\gamma} \right] \frac{1+(1-y)^2 -y^2 
\left[ 1 - {r^2 \over 1 + R^{\gamma Z}} \right] - 2 x y {M\over E}} {1+(1-y)^2 -y^2 
\left[1- {r^2 \over 1+R^{\gamma}} \right] - 2 x y  {M\over E}}\ , \nn \\
&& Y_3 = \left[ \frac{r^2}{1+R^\gamma} \right] \frac{1-(1-y)^2 }{1+(1-y)^2 -y^2 
\left[ 1 - {r^2 \over 1 + R^{\gamma}} \right] - 2 x y {M\over E}}\ , 
\eea
where $R^\gamma$ and $R^{\gamma Z}$ denote the ratios of the longitudinal to transverse virtual photon cross-sections, 
for the EM ($\gamma$) and the interference ($\gamma Z$) contributions, respectively. 
They are given in terms of the structure functions as
\be
\label{apv15}
R^{\gamma (\gamma Z)} = r^2 \frac{F_2^{\gamma (\gamma Z)}}{2xF_1^{\gamma (\gamma Z)} }-1\ , \qquad\qquad
r^2 \equiv 1 + \frac{4M^2x^2}{Q^2}\ , \qquad\qquad 
x \equiv {Q^2 \over y s}\ .
\ee
In the (Bjorken) limit, $Q^2 \to \infty$, in which the Bjorken variable $x$ is kept fixed, 
the structure functions satisfy the Callan-Gross relations, $F_2^{\gamma (\gamma Z)}=2xF_1^{\gamma(\gamma Z)}$, and
\be
Y_1 \to 1\ , \qquad\qquad Y_3 \to \frac{1-(1-y)^2}{1+ (1-y)^2}\ .
\ee
In this limit and ignoring sea quarks and CSV, $A_{LR}^{\rm DIS}$ for the isoscalar deuteron target 
takes the simple form in Equation~(\ref{eDIS}).
All structure function effects cancel, allowing for a clean extraction of the $g^{eq}_{AV}$ and $g^{eq}_{VA}$ coefficients.

\subsection{Higher Twist Effects}

The PVDIS asymmetry for electron-deuteron scattering is particularly interesting 
as a probe of long range quark and gluon correlations,
which go beyond the leading twist (twist-2) parton model.
They give rise to higher twist $Q^2$-dependent power corrections encoded in $R_{1}({\rm HT})$ and $R_2({\rm HT})$.  
It was first shown in References~\cite{Bjorken:1978ry,Wolfenstein:1978rr} that the twist-4 contribution 
to $R_{1}({\rm HT})$ is due to the deuteron ($D$) matrix element of a single four-quark operator,
\be
\label{twist4}
W^{du}_{\mu \nu} = \frac{1}{M_D} \int {d^4x \over 2\pi}\, e^{i q\cdot x}
\langle D(\vec P) | {\bar{d}(x)\gamma_\mu d(x)\, \bar{u}(0)\gamma_\nu u(0) + (u\leftrightarrow d)\over 2} |D(\vec P)\rangle\ .
\ee
$R_{2}({\rm HT})$, which receives contributions from several twist-4 operators, is relatively suppressed 
by $g^{eq}_{VA}$ and can be isolated from $R_{1}({\rm HT})$ through its $y$-dependence.  
It was also shown in References~\cite{Bjorken:1978ry,Wolfenstein:1978rr} that $R_{2}({\rm HT})$ 
is related to the higher twist effects appearing in neutrino-deuteron charged-current DIS. 
These properties allow for a relatively clean theoretical interpretation of the quark-quark correlation in Equation~\ref{twist4}, 
provided that $R_{1}({\rm HT})$ is large enough to be observed. 

Several earlier 
works~\cite{Fajfer:1984um,Castorina:1985uw,Dasgupta:1996hh,Stein:1996wk,Signal:1996ct,Stein:1998wr,Beneke:1998ui} 
have explored various aspects of HT effects in PVDIS.
More recently~\cite{Hobbs:2008mm,Mantry:2010ki}, the HT phenomenology was revisited 
in the context of the current generation of high precision experiments. 
In the form of $A_{LR}^{\rm DIS}$ given in Equation~\ref{apv1}, the $Y_1$ term can receive contributions 
from HT effects if $R^{\gamma Z}\neq R^\gamma$ and from the ratio $F_{1}^{\gamma Z}/F_{1}^\gamma$. 
In the absence of empirical data on $R^{\gamma Z}$, the impact of $R^{\gamma Z}\neq R^{\gamma}$ 
was explored in Reference~\cite{Hobbs:2008mm} and it was found that the variation 
$R^{\gamma Z}= R^\gamma \pm 10\%$ resulted in a $\sim 0.5\%$ shift in $A_{LR}^{\rm DIS}$. 
In subsequent work~\cite{Mantry:2010ki}, the Bjorken-Wolfenstein~\cite{Bjorken:1978ry,Wolfenstein:1978rr} 
argument was applied to the $Y_1$ term and it was shown that the equality $R^{\gamma Z} = R^{\gamma}$ 
was true even at twist-4 up to perturbative corrections ---
a consequence of the twist-4 structure functions associated with $W_{\mu \nu}^{du}$ satisfying the tree-level 
Callan-Gross relation $F_2^{du}=2x F_1^{du}$~\cite{Ellis:1982cd,Ji:1993ey,Qiu:1988dn}. 
Instead, the dominant HT effects arises through the ratio $F_{1}^{\gamma Z}/F_{1}^\gamma$, 
giving rise to the correction
\be
R_1({\rm HT}) = \frac{-4}{5 \left( 1 - \frac{20}{9}\sin^2\theta_W \right) }\frac{F_1^{du}}{u_p(x)+d_p(x)}\ ,
\ee
where $u_p$ ($d_p$) is the $u$ ($d$) quark parton distribution function (PDF) in the proton.

The first estimates of the twist-4 contribution to $R_1({\rm HT})$ were obtained~\cite{Fajfer:1984um,Castorina:1985uw} 
within the MIT Bag Model~\cite{Chodos:1974je} by rescaling the twist-2 contribution by the ratio of their leading moments. 
This  computation was recently~\cite{Mantry:2010ki} extended to include the effects of higher spin operators. 
A model for the nucleon wave functions in the light-cone formalism~\cite{Bolz:1996sw,Diehl:1998kh,Braun:2011aw} 
was used to obtain an independent estimate~\cite{Belitsky:2011gz} for $R_1({\rm HT})$. 
Most recently~\cite{Seng:2013fia}, quark orbital angular momentum dynamics was added 
and its effect on $R_1({\rm HT})$ was studied. 
It was found that $R_1({\rm HT})$ was largely insensitive to quark angular momentum 
due to cancellations between its different components. 
All of these recent studies yield similar results in the range $R_1({\rm HT}) \sim 0.002$ to 0.005 in the valence region of $x$. 
This is likely too small to be measured by SoLID which is expected to measure $A_{LR}^{\rm DIS}$ to $\sim 0.5\%$ precision, 
and indicates that $A_{LR}^{\rm DIS}$ can probe CSV or new physics without HT contamination. 
On the other hand,  if a large $Q^2$-dependent effect is observed in the $Y_1$-term, 
it would be a clear indication of interesting long range quark-quark correlations 
that are not adequately described by current theoretical models.

\subsection{Charge Symmetry Violation}

Most early phenomenological work on PDF assumed charge symmetry,
meaning that the $u$ ($d$) quark PDF in the proton was set equal to the $d$ $(u)$ quark PDF in the neutron
\be
u_p(x) = d_n(x)\ , \qquad\qquad 
d_p(x)=u_n(x)\ .
\ee 
CSV is expected to arise from quark mass differences and the effects 
of QED splitting functions~\cite{Martin:2003sk,Martin:2004dh,Gluck:2005xh} on the DGLAP evolution of PDF. 
Several non-perturbative models~\cite{Sather:1991je,Londergan:1994gr,Boros:1999fy} of the nucleon 
have been studied to estimate the size of CSV effects. 
While no conclusive evidence of CSV has been observed, 
it is constrained from an analysis of high energy data~\cite{Martin:2003sk}. 
The strongest limit on CSV arises from comparison of the $F_2$ structure functions 
between charged lepton DIS and neutrino charged current DIS on isoscalar nuclear targets. 
For a recent detailed review on CSV and experimental constraints, we refer the reader to Reference~\cite{Londergan:2009kj}.
 
The CSV effects can be parameterized in terms of the quantities $\delta u$ and $\delta d$, 
\be
\delta u(x) \equiv u_p(x)-d_n(x)\ , \qquad\qquad 
\delta d(x) \equiv d_p(x)-u_n(x)\ .
\ee 
Then, the CSV corrections to $A_{LR}^{\rm DIS}$ in Equations~\ref{apv2} and \ref{apv3} take the form,
\be
\label{Rcsv}
R_{i}({\rm CSV}) = \left[ \frac{1}{2} \left( \frac{2 g^{eu}_{i} + g^{ed}_{i}}{2 g^{eu}_{i} - g^{ed}_{i}} \right) - \frac{3}{10} \right] 
\left( \frac{\delta u - \delta d}{u+d} \right)\ .
\ee

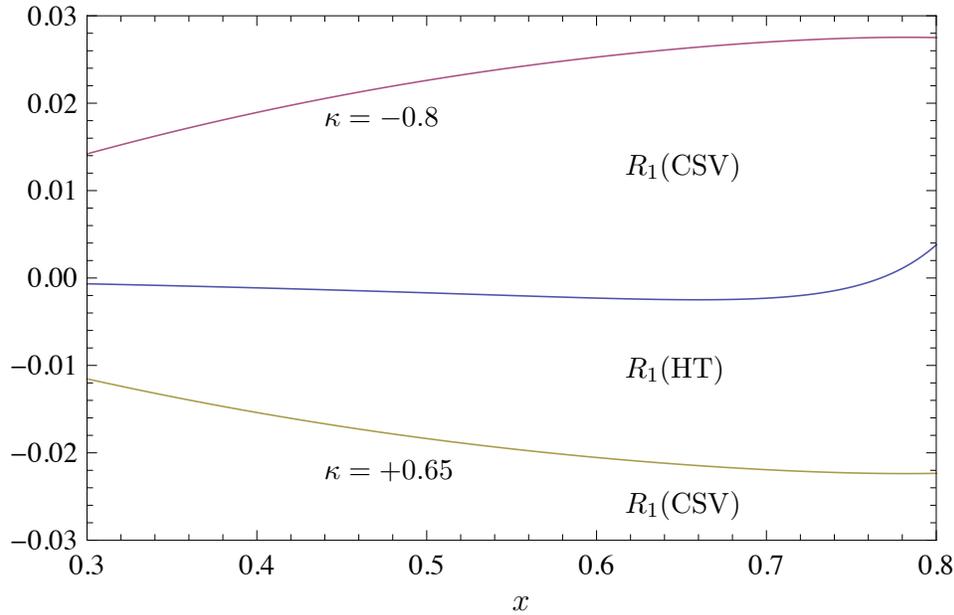
\begin{figure}[!t]
\begin{center}
\includegraphics[width=0.93\textwidth]{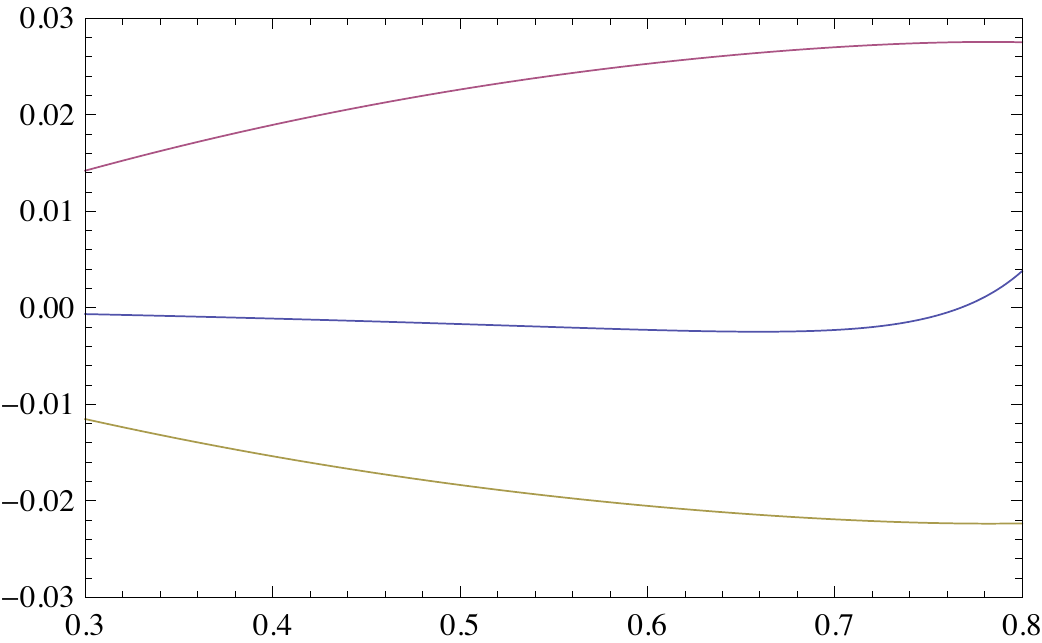}
\rput[tl](-4.5,5.7){$R_1({\rm CSV})$}
\rput[tl](-4.5,3.0){$R_1({\rm HT})$}
\rput[tl](-4.5,1.2){$R_1({\rm CSV})$}
\rput[tl](-6, -0.2){\large{$x$}}
\rput[tl](-8.5,6.3){$\kappa=-0.8$}
\rput[tl](-8.5,1.6){$\kappa=+0.65$}
\caption{The relative magnitudes of $R_1({\rm HT})$ and $R_1({\rm CSV})$ as a function of $x$ 
for a representative value of $Q^2 = 6$~GeV$^2$. 
The top and bottom curves give $R_1({\rm CSV})$ for the choices $\kappa=-0.8$ and $\kappa=0.65$,
respectively, using Equations~\ref{Rcsv} and \ref{dudd}. 
The middle curve~\cite{Mantry:2010ki} is the MIT Bag Model estimate for $R_1({\rm HT})$.
Figure reprinted from Reference~\cite{Mantry:2010ki}.}
\label{CSV}
\end{center}
\end{figure}

A global fit to high energy data was performed by the MRST group~\cite{Martin:2003sk} using the phenomenological form,
\be
\label{dudd}
\delta u - \delta d = 2 \kappa x^{-1/2} (1-x)^4(x-0.0909)\ ,
\ee
that yielded a best fit value of $\kappa = -0.2$ and a $90\%$ confidence interval given by $-0.8 \leq \kappa \leq 0.65$. 
In {\bf Figure~\ref{CSV}}, we show the CSV correction $R_1({\rm CSV})$ for the end points of this interval.
For comparison we also show the MIT Bag model estimate~\cite{Mantry:2010ki} of the HT correction $R_1({\rm HT})$. 
This estimate for $R_1({\rm HT})$, similar in size to other 
estimates~\cite{Fajfer:1984um,Castorina:1985uw,Belitsky:2011gz,Seng:2013fia}, indicates that the $Y_1$-term 
of $A_{LR}^{\rm DIS}$ can probe percent level CSV effects, without much contamination from HT effects. 
The CSV effects on the total asymmetry, including both $R_1({\rm CSV})$ and $R_2({\rm CSV})$,
are also at the percent level~\cite{Hobbs:2008mm} of the $90\%$ confidence interval, 
which should be contrasted with the expected precision of SOLID at the $0.5\%$ level.

Flavor-dependent effects can also arise for PDF in heavier nuclei 
through isospin-dependent nuclear forces affecting the $u$- and $d$ quark distributions differently. Several 
works~\cite{LlewellynSmith:1983qa,Bickerstaff:1989ch,Ericson:1983um,Jaffe:1982rr,Alde:1990im,Saito:1992rm,Weinstein:2010rt} 
have been devoted to understanding the underlying mechanisms 
responsible for differences in the cross sections per nucleon among various nuclei. 
Recently, nuclear-dependent effects were  also studied~\cite{Kang:2013wca} using a newly 
introduced~\cite{Kang:2012zr,Kang:2013wca,Kang:2013nha,Kang:2013lga} DIS event shape called 1-jettiness~\cite{Stewart:2010tn}.  
Differences in the PDF in the valence region of $x$, due to nuclear effects, are referred to as the EMC~\cite{Aubert:1983xm} effect. 
An unambiguous picture of the mechanisms responsible for the EMC effect is still lacking after more than two decades of investigation.

Recently, PVDIS on nuclei has been proposed~\cite{Cloet:2012td} 
as a novel probe of the flavor dependence of the nuclear EMC effect,
which appears dominantly as an isovector correction to the  $\tilde{a}_1$ term in Equation~\ref{apv2},
\be
\delta \tilde{a}_1(x_A) \simeq  -\frac{12}{25}\frac{u_A^+(x_A) -d_A^+(x)}{u_A^+(x_A) +d_A^+(x)}\ ,
\ee
where $x_A$ is the parton momentum fraction in the nucleus multiplied by the atomic weight $A$ and 
$q_A^+(x_A) \equiv q_A(x_A)+\bar{q}_A(x_A)$. 
For isoscalar targets, ignoring heavy quark flavors, quark mass differences, and EW corrections, 
this correction vanishes since the $u$ and $d$ quark nuclear distributions $u_A$ and $d_A$ are identical. 
However, for other nuclear targets with $N\neq Z$, this isovector EMC correction could be large. 
Such a PVDIS analysis on an iron or lead target could provide insight into the impact of this EMC effect 
on the extraction of the weak mixing angle at NuTeV~\cite{Aubert:1983xm}. 
In general, combining $\delta \tilde{a}_1(x_A)$ from PVDIS analyses with data on EM DIS,  
may allow to extract flavor-dependent nuclear PDF.

\subsection{Extracting the $d/u$ Ratio of Parton Distribution Functions}

A measurement of $A_{LR}^{\rm DIS}$ on a proton target 
is sensitive~\cite{Souder:2005tz} to the ratio of the $d$ to $u$ quark PDF. 
The standard determination of the $d/u$ ratio relies on fully inclusive DIS on a proton target compared to a deuteron target. 
In the large $x$ region, nuclear corrections in the deuteron target lead to large uncertainties in the $d/u$ ratio. 
Several methods~\cite{Saito:2000fx,Afnan:2003vh,Frankfurt:1981mk,Simula:1996xk,Melnitchouk:1996vp,Melnitchouk:1999un} 
to control these nuclear uncertainties have been investigated.  
However, they can be completely eliminated if the $d/u$ ratio is obtained from the proton target alone. 
For this reason, precision measurements of $A_{LR}^{\rm DIS}$ on a proton target can be a powerful probe of the $d/u$ ratio.

$A_{LR}^{\rm DIS}$ in Equation~\ref{apv1} for a proton target at leading twist takes the form,
\be
A_{\rm LR}^p = - \frac{1}{4\pi \alpha} \frac{Q^2}{v^2}\,\left[ Y_1\, a_1^p + Y_3\, a_2^p\,\right]\ ,
\ee
where the coefficients $a_1^p$ and $a_3^p$ depend on the $d/u$ ratio,
\be
a_1^p = \frac{12\, g^{eu}_{AV} - 6\, g^{ed}_{AV} \>d/u}{4+ d/u}\ , \qquad\qquad 
a_2^p = \frac{12 \, g^{eu}_{VA}-6\> g^{ed}_{VA} \, d/u}{4+d/u}\ .
\ee
As in the case of the deuteron asymmetry, other hadronic corrections can affect the extraction of the $d/u$ ratio. 
Reference~\cite{Hobbs:2008mm} performed an analysis of finite-$Q^2$ effects in the $Y_1$ and $Y_3$ factors 
and studied their impact on the extraction of $d/u$. 
For finite-$Q^2$ effects arising from  $r\neq 1$ and $R^\gamma\neq 0$, where the parameterization 
in Reference~\cite{Whitlow:1990gk} was used for $R^\gamma$, a shift in $A_{LR}^{\rm DIS}$ in the 1~to~2\% range was found
at $Q^2= 5$~GeV$^2$ and for $0.6 \lesssim x \lesssim 0.8$, with an uncertainty of $\pm 0.5\%$.  
This shift increased to about $3$\% for $x\simeq 0.9$ with an uncertainty of $\pm 1 \%$. 
Another hadronic effect arises from possible differences between $R^\gamma$ and $R^{\gamma Z}$. 
It was found~\cite{Hobbs:2008mm} that a 10\% (20\%) difference led to a 1\% (2\%) shift in $A_{LR}^{\rm DIS}$.
Both finite-$Q^2$ effects are to be compared with shifts in the 3~to~10\% range arising from different possible 
behaviors~\cite{Tung:2006tb,Melnitchouk:1995fc,Melnitchouk:1996fh} of the $d/u$ ratio at large $x$.  
This analysis indicates that precision measurements of PVDIS on a proton target could provide 
useful information on the $d/u$ ratio in the region of large $x$.

\section{NUCLEAR PHYSICS}
\label{secnuclear}


PVES can map the distribution of weak charge in a nucleus.  
This provides largely model independent neutron densities 
because the weak charge of a neutron is much larger than that of a proton.   
In this section we discuss the PREX and PREX~II experiments on $^{208}$Pb 
and the approved CREX experiment on $^{48}$Ca.  
We conclude this section commenting on a precision measurement of the weak charge of $^{12}$C 
from electron scattering at low $Q^2$.
This would involve precision comparable to or better than the approximately 0.3\% APV measurement of the weak charge of Cs.   
One would need high statistics, small systematic errors and accurate normalization, 
including high accuracy measurements of beam polarization.  
However, if this could be achieved the resulting determination of the weak charge of $^{12}$C 
would be free from atomic structure uncertainties that may complicate the interpretation of APV.

The formalism is presented in Section~\ref{subsec.nuc.formalism}, 
while Section~\ref{subsec.nuc.symmetryE} relates the atomic PV asymmetry 
to the thickness of the expected neutron rich skin and to the density dependence of the symmetry energy.  
This important nuclear structure quantity describes how the energy of nuclear matter increases 
as one goes away from equal numbers of neutrons and protons.  
Section~\ref{subsec.nuc.astrophysics} discusses important applications to astrophysics 
for inferring the pressure of neutron matter and the structure of neutron stars.  
Finally, in Section~\ref{subsec.nuc.exp} we briefly discuss 
the experiments on $^{208}$Pb (PREX and PREX~II), $^{48}$Ca (CREX), and $^{12}$C.


\subsection{Nuclear Formalism}
\label{subsec.nuc.formalism}
In this section we calculate the weak charge density of a heavy nucleus and the PV asymmetry $A_{LR}$.  
For simplicity we consider spin zero nuclei.  
We start by defining the weak charges of the proton $Q_p$ and neutron $Q_n$ as 
minus twice the corresponding coupling constants of {\bf Table~\ref{ge}}.  
We use $Q_p=-2g_{AV}^{\ e p} = 0.0710$ and $Q_n=-2 g_{AV}^{\ e n} = -0.9902$, 
and recall that $Q_p$ is small, includes important radiative corrections, 
and is being measured by the Qweak experiment~\cite{Androic:2013rhu}.  
In contrast, $Q_n$ is large, includes modest radiative corrections, and is constrained by Cs APV.
We reiterate (see Section~\ref{radcorr}) that at the level of small radiative corrections these definitions 
of $Q_p$ and $Q_n$ differ from the standard convention~\cite{Marciano:1982mm}.

The weak charge density of a heavy nucleus,
\be
\rho_W(r)=\int d^3r'[4G_n^Z(|{\bf r}-{\bf r'}|)\rho_n(r')+4G_p^Z(|{\bf r}-{\bf r'}|) \rho_p(r')]\ ,
\ee     
is modeled as point proton and neutron densities, $\rho_p$ and $\rho_n$, folded with appropriate single nucleon weak form factors.  
The densities $\rho_p$ and $\rho_n$ are normalized to the proton, $Z=\int d^3r\rho_p(r)$,
and neutron, $N=\int d^3r\rho_n(r)$, numbers of the nucleus. 
This neglects possible meson exchange currents that are expected to be small because mesons likely transport 
weak charge only over distances small compared to the nuclear weak radius~\cite{PhysRevC.63.025501}.  
However, see also the spin-orbit contributions to weak charge densities discussed in~\cite{PhysRevC.86.045503}.

The weak single nucleon form factors of the proton, $G_p^Z$, and neutron, $G_n^Z$, 
Fourier transformed into coordinate space, are
\bea
4G_p^Z=Q_p G_E^p+Q_nG_E^n-G_E^s\ , \\
4G_n^Z =Q_nG_E^p+Q_p G_E^n -G_E^s\ .
\eea
Here the (EM) Electric form factor of the proton is $G_E^p$ and the Electric form factor of the neutron is $G_E^n$.  
Finally, strange quark contributions to the Electric form factor, $G_E^s$, are constrained 
by several previous measurements and will be neglected in the following.  
The weak form factor $F_W(Q^2)$ is defined,
\be
F_W(Q^2)=\frac{1}{Q_W}\int d^3r j_0(Qr) \rho_W(r)\ ,
\ee
where $j_0$ is a spherical Bessel function.
The total nuclear weak charge is given by
\be
Q_W=\int d^3r \rho_W(r)=NQ_n+ZQ_p\ .
\ee

In the Born approximation, the PV cross-section asymmetry for longitudinally polarized electrons elastically scattered 
from an unpolarized nucleus,  $A_{LR}$, is
\begin{equation}
A_{LR} \approx \frac{1}{8\pi \alpha}\, \frac{Q^2}{v^2}\, \frac{Q_W}{Z}\, \frac{F_W(Q^2)}{F_{ch}(Q^2)}\ ,
\label{born_asy}
\end{equation}
where $F_{ch}(Q^2)$ is the Fourier transform of the known charge density and is normalized so that $F_{ch}(Q^2=0)=1$.  

For a heavy nucleus, the Born approximation is not adequate and 
there are important corrections from Coulomb distortions that are of order $Z\alpha$.  
It is useful to distinguish Coulomb distortions, that involve the exchange of an additional photon 
where the nucleus remains in the same state, from dispersion corrections.  
Dispersion corrections involve excited intermediate states and are order $\alpha$ (instead of $Z\alpha$).  
Comparison of $^{208}$Pb cross section measurements with positrons and electrons 
suggest that dispersion corrections are small~\cite{PhysRevLett.66.572}.

Coulomb distortions can be accurately calculated by solving the Dirac equation 
for an electron moving in both a Coulomb potential, that is of order 25 MeV for $^{208}$Pb, 
and an axial vector potential of order $G_F\rho_W\approx 10$ eV~\cite{PhysRevC.57.3430}.  
The cross section for positive helicity involves an electron scattering in the sum of axial and vector potentials, 
while the cross section for negative helicity involves scattering from the difference of the vector minus the axial potentials.  
These Coulomb distortion calculations are good to all orders in $Z\alpha$ 
and involve little uncertainty because the charge density is accurately known.  


\subsection{Neutron Skins and the Symmetry Energy}
\label{subsec.nuc.symmetryE}

PVES can determine the neutron skin thickness, $\Delta R=R_n-R_p$, 
defined as the difference between the root mean square (point) neutron radius, $R_n$, and the proton radius, $R_p$.  
Note that $R_p$ is often known from measured charge radii.  
Theoretical predictions for $\Delta R$, for nuclei all across the periodic table, 
are discussed in Reference~\cite{2013PhRvC..88c1305K}.  
The skin thickness is determined largely by isovector parts of nuclear interactions that are, in general, 
not well constrained by fitting binding energies and charge radii of conventional nuclei.

Instead, the skin thickness is closely related to the density dependence of the symmetry energy.  
The energy per particle of asymmetric matter, $\rho_n \neq \rho_p$, is
\be 
\frac{E}{A}(\rho_n\neq\rho_p)\approx \frac{E}{A}(\rho_n=\rho_p)+ \alpha^2 S(\rho)\ .
\ee
Here the neutron excess is $\alpha=[(\rho_n-\rho_p)/\rho]^2$ with $\rho=\rho_n+\rho_p$.  
The symmetry energy $S(\rho)$ describes how the energy of nuclear matter increases 
as one moves away from equal numbers of neutrons and protons.  
It is a function of density $\rho$, and arrises from the Pauli exclusion principle 
and because nucleon-nucleon interactions are more attractive in isospin zero, compared to isospin one states. 

As an example, $^{208}$Pb has 44 more neutrons than protons.  
If these extra neutrons are placed at a density $\rho$ in the nucleus, than there will be an energy cost $S(\rho)$.  
If the symmetry energy is independent of density, surface tension will push the extra neutrons to high densities 
and the neutron skin will be small.  
However, if $S\rho)$ increases rapidly with density, the symmetry energy will favor 
putting the extra neutrons at low densities in the nuclear surface and this will give a large $\Delta R$.  
Therefore there is a strong correlation between the neutron skin thickness in $^{208}$Pb 
and the density dependence of the symmetry energy $L$~\cite{PhysRevLett.106.252501}.
This parameter is defined as,
\be
L=3\rho_0 \frac{dS}{d\rho}\Bigr|_{\rho_0}\ ,
\ee
where the nuclear saturation density is $\rho_0=0.16$ fm$^{-3}$.   
There is a great deal of interest in determining $L$ from other nuclear structure measurements 
and heavy ion collisions~\cite{PhysRevC.86.015803}.  
Measuring $\Delta R$ with PVES likely provides the most model independent way to determine $L$.

\subsection{Neutron Skins and Astrophysics}
\label{subsec.nuc.astrophysics}

The neutron skin thickness is also important for astrophysics.  
The pressure of neutron matter is closely related to the density dependence of the symmetry energy.   
Typel and Brown showed that there is a strong correlation between the neutron skin thickness in $^{208}$Pb, 
as predicted by many density functionals, 
and the pressure of neutron matter at a density near 0.1 fm$^{-3}$~\cite{PhysRevC.64.027302}.  
The larger the pressure, the more neutrons are pushed out against surface tension, and the larger the neutron skin.

The structure of neutron stars is determined by the equation of state, pressure versus energy density, of neutron rich matter.  
The higher the pressure, the further out matter is supported against gravity and the larger the neutron star radius.  
Thus, in general one expects a correlation between the neutron skin thickness in $^{208}$Pb and 
the radius of a neutron star~\cite{PhysRevC.64.062802,PhysRevLett.86.5647}.  
A thick neutron skin suggests a large pressure and a large neutron star radius.

There is great interest in X-ray observations of neutron star radii $R_{\rm NS}$.  
If one can determine the luminosity $L_{X-ray}$ and surface temperature $T$, from X-ray spectra, 
one can infer an effective surface area $4\pi R_{\rm NS}^2$,
\be
L_{X-ray}=4\pi R_{\rm NS}^2 \sigma T^4\ ,
\label{Eq.RNS}
\ee
where $\sigma$ is the Stephan Boltzmann constant.  
However, there are important complications.  
One needs an accurate distance to the star in order to determine the absolute luminosity $L_{X-ray}$.  
E.g., one can identify a neutron star (NS) as a member of a globular star cluster of known distance.  
Equation~\ref{Eq.RNS} assumes a black body.  
There are important corrections to this from atmosphere models that may depend on composition 
and magnetic field~\cite{2012A&A...545A.120S}.  
Finally, gravity is so strong that space is curved near a NS.  
If one observes the near face of a NS one also sees about 30\% of the far face as light is bent around the star.  
Therefore, the effective surface area $4\pi R_{\rm NS}^2$ actually depends on a mixture of radius 
and mass of the star (because the curvature of space depends on the mass).   
X-ray observations of NS radii by Ozel et al.~\cite{2013RPPh...76a6901O}, Guillot et al.~\cite{2013ApJ...772....7G}, 
Steiner et al.~\cite{2013arXiv1305.3242L}, and Suleimanov et al.~\cite{2011ApJ...742..122S} 
yield a range of radii from $R_{\rm NS}\approx 10$ km to more than $14$~km.  
E.g., Steiner et al.\ infer $R_{\rm NS}\approx 12$ km and, from the equation of state that they deduce from 
X-ray observations of NS, the neutron skin thickness in $^{208}$Pb should be $<0.2$ fm.  
The Large Observatory for Timing (LOFT) is a proposed European Space Agency mission 
that should improve our knowledge of $R_{\rm NS}$~\cite{2013arXiv1312.1697B}.

The neutron skin thickness in $^{208}$Pb is also important for other properties of NS in addition to radii.  
E.g., the crust core transition density in a NS is correlated with $\Delta R$~\cite{PhysRevLett.86.5647}.  
Finally, the interior composition and neutrino emissivity of a NS depends on the density dependence 
of the symmetry energy and $\Delta R$~\cite{PhysRevC.66.055803}.


\subsection{Nuclear Experiments}
\label{subsec.nuc.exp}

In this subsection we briefly describe the nuclear experiments on $^{208}$Pb (PREX and PREX~II), 
$^{48}$Ca (CREX), and on $^{12}$C (see {\bf Table~\ref{tab:expts:A}}).  
PREX involved the scattering of 1~GeV electrons at about five degrees from a 0.5 mm thick 
$^{208}$Pb target~\cite{Abrahamyan:2012gp}.  
The measured asymmetry,
\be
A_{LR}= - 0.656\pm0.060~({\rm stat})\pm 0.014~({\rm syst})~{\rm ppm}\ ,
\ee
involved a relatively small systematic error and a larger statistical error.   
From this measurement the weak form factor was deduced to be~\cite{PhysRevC.85.032501},
\be
F_W(Q^2=0.0088\ {\rm GeV}^2)=0.204\pm 0.028\ ,
\ee
and the root mean square radius of the weak charge distribution,
\be
R_W=5.826\pm 0.181\ {\rm fm}\ .
\ee  
This can be directly compared to the well known (EM) charge radius of $^{208}$Pb of 5.503 fm.  
This shows that the weak charge distribution of a heavy nucleus is more extended than the EM charge distribution.  
This is closely related to the expected neutron rich skin.  
By comparing predictions of $A_{LR}$ for a range of relativistic mean field models, 
the neutron skin thickness is deduced to be~\cite{Abrahamyan:2012gp},
\be
\Delta R=R_n-R_p= 0.33_{-0.18}^{+0.16}\ {\rm fm}\ .
\ee
The error in $\Delta R$ is dominated by the statistical error in $A_{LR}$.   
This relatively large value for $\Delta R$ suggests that neutron stars should be relatively large $R_{\rm NS}>12$ km.  
However, the error bars are large.  
The goal of the approved PREX~II experiment is to perform a second measurement of $^{208}$Pb, 
at the same kinematics as PREX, in order to improve the statistical error on $A_{LR}$ 
and deduce $\Delta R$ with a three times smaller error of $\pm 0.06$ fm.

The CREX experiment aims to measure $\Delta R$ for $^{48}$Ca to $\pm0.02$~fm~\cite{CREX}.  
Like $^{208}$Pb, $^{48}$Ca is also a neutron rich nucleus with both closed proton and neutron shells (doubly magic).  
However $^{48}$Ca is significantly lighter than $^{208}$Pb.  
This allows microscopic coupled cluster calculations for $^{48}$Ca~\cite{PhysRevLett.109.032502} 
that are presently not feasible for $^{208}$Pb.  
These calculations can relate $\Delta R$ to very interesting three neutron forces.  
Measuring $\Delta R$ for both the heavy $^{208}$Pb and light $^{48}$Ca will allow one to constrain 
isovector terms in the energy functional describing both surface (gradient) and volume energies.

Finally, a precision measurement of $A_{LR}$ for $^{12}$C 
can accurately constrain the weak charge of $^{12}$C~\cite{2013arXiv1311.1843M}.  
The weak charge of $^{12}$C involves the isoscalar combination of proton and neutron weak charges and 
constrains new physics in a way similar to the weak charge of Cs (as determined from APV).  
Because the atomic number of carbon is only 6, Coulomb distortion effects are small and can be accurately calculated.  
Furthermore, other nuclear structure corrections such as $\Delta R$ are also expected to be small at low momentum transfers.  
Therefore, a $^{12}$C measurement should allow one to constrain the weak charge of a nucleus 
in a way that is free from atomic structure uncertainties that are important for APV.

To summarize, PVES from a nucleus can constrain the neutron skin thickness, $\Delta R=R_n-R_p$,
and the density dependence of the symmetry energy.  
This is important for nuclear structure.  
Furthermore, $\Delta R$ is important in astrophysics to constrain neutron star radii.  
Finally, both PVES and APV can constrain the weak charge of a nucleus.

\section{PROBING NEW PHYSICS}
\label{secnewphysics}


PVES and related experiments can be used to search for evidence for new particles
or interactions beyond those of the SM, or conversely to set exclusion limits.
For example, new particles may modify the self-energies of the electroweak gauge
bosons, leading to observable shifts in fundamental parameters such as $\sstw$.
This kind of effect is described by what are known as ``oblique parameters", 
and has been reviewed very recently in References~\cite{Kumar:2013yoa,Erler:2013xha}.
But the new physics may also generate new quantum-mechanical amplitudes.
These and other phenomena may be correlated in specific scenarios and their
realizations in concrete models.


\subsection{Beyond the SM Amplitudes and New Physics Scales}

The NC Lagrangian in Equation~\ref{LeqNC} shows four-Fermi contact interactions
normalized with respect to the EW scale, $v$. 
In the presence of non-standard physics there will be new contributions, so that e.g.,
\begin{equation}
\label{Lnew}
\frac{g_{AV}^{eq}}{2 v^2} \bar{e} \gamma^\mu  \gamma^5 e \bar{q}\, \gamma_\mu q \to
\left[ \frac{g_{AV}^{eq}}{2 v^2} + \frac{4\pi}{({\Lambda^{eq}_{AV}})^2} \right]
\bar{e} \gamma^\mu  \gamma^5 e \bar{q}\, \gamma_\mu q\ ,
\end{equation}
and similarly for the other interaction types.
In particular, new contact interactions are expected if leptons or quarks have a substructure and are composed 
of more fundamental objects, bound together by a new interaction of non-perturbative strength.
Therefore, it has become conventional~\cite{Eichten:1983hw} to choose the new coefficients in~\ref{Lnew} 
equal to $4\pi$, and to parametrize the effects of the new operators in terms of compositeness scales $\Lambda$,
but the resulting bounds can be rescaled to apply to more general scenarios of new physics with strength
$g_{\rm new} \neq 4\pi$.

As an example, if one considers only models with positive- or negative-definite contributions
(sometimes this sort of assumption is itself referred to as a model)
then the constraint~\ref{SLAC-E158} implies, respectively, the 95\% CL upper and lower limits,
\begin{equation}
|g_{VA}^{ee}|^+ = 0.0035\ , \qquad\qquad
|g_{VA}^{ee}|^- = 0.0081\ ,
\end{equation}
and one obtains,
\begin{equation}
\label{E158+}
\Lambda_+ > v \sqrt{\frac{8\pi}{|g_{VA}^{ee}|^+}} = 20.9~{\rm TeV}\ , \qquad\qquad
\Lambda_- > v \sqrt{\frac{8\pi}{|g_{VA}^{ee}|^-}} = 13.7~{\rm TeV}\ .
\end{equation}
The MOLLER experiment is expected to achieve $\Delta g_{VA}^{ee} = 0.00052$ (2.3\%) so that
\begin{equation}
\label{MOLLERexpected}
\Lambda_\pm ({\rm expected}) > v \sqrt{\frac{8\pi}{1.96\, \Delta g_{VA}^{ee}}} = 38.7~{\rm TeV}.
\end{equation}

\begin{table}[!t]
\caption{Achieved (upper panel) or anticipated (lower panel) relative uncertainty, 
the corresponding tree level sensitivity to extract the weak mixing angle, and 
the expected reach to the associated compositeness scales (at 95\% CL) for key parity violation experiments.
For hadronic probes we also give the relevant linear combinations as the angle $\theta$ relative to
$2\, g_{AV}^{eu} - g_{AV}^{ed}$.}
\begin{center}
\begin{tabular}{l|r|l|r|r}
\hline
 & precision~(\%) & $\Delta\sin^2\hat\theta_W(0)$ & $\Lambda_{\rm new}$~[TeV] & $\theta$ \\
\hline
SLAC--E122           &   8.3\ph{000} & \ph{00}0.011 & \ph{1}5.3\ph{00} & $9.4^\circ$ \\
SLAC--E122           &  110\ph{.0000} & \ph{00}0.44 & \ph{1}0.9\ph{00}  & $99.4^\circ$ \\
APV ($^{205}$Tl)      &   3.2\ph{000} & \ph{00}0.011 & 13.5\ph{00} & $75.6^\circ$ \\
APV ($^{133}$Cs)    & 0.58\ph{00} & \ph{00}0.0019 & 32.3\ph{00} & $74.9^\circ$  \\
SLAC--E158           &    14\ph{.0000} & \ph{00}0.0013   & 17.0\ph{00}  & --- \ph{$^\circ$} \\
JLab--Qweak (run I) &    19\ph{.0000} & \ph{00}0.0030   & 17.0\ph{00} & $53.1^\circ$ \\
JLab--Hall A              &   4.1\ph{000} & \ph{00}0.0051     & \ph{1}7.8\ph{00}  & $26.2^\circ$ \\
JLab--Hall A              &    61\ph{.0000} & \ph{00}0.051     & \ph{1}2.9\ph{00}  & $ \ph{0}116.2^\circ$ \\
\hline
JLab--Qweak (final) &  4.5\ph{000} & \ph{00}0.0008   & 33\ph{.000} & $53.1^\circ$ \\  
JLab--SoLID             &  0.6\ph{000} & \ph{00}0.00057 & 22\ph{.000}  & $40.0^\circ$ \\
JLab--MOLLER        &  2.3\ph{000} & \ph{00}0.00026 & 39\ph{.000}  & --- \ph{$^\circ$}  \\
Mainz--P2                  &  2.0\ph{000} & \ph{00}0.00036 & 49\ph{.000}  & $53.1^\circ$ \\
APV ($^{225}$Ra$^+$) & 0.5\ph{000} & \ph{00}0.0018 & 34\ph{.000} & $75.7^\circ$ \\
APV $(^{213}$Ra$^+/^{225}$Ra$^+$) & 0.1\ph{000} & \ph{00}0.0037 & 16\ph{.000} & $55.5^\circ$ \\
PVES ($^{12}$C)     &  0.3\ph{000} & \ph{00}0.0007 & 49\ph{.000} & $71.6^\circ$ \\
\hline
\end{tabular}
\label{reach}
\end{center}
\end{table}

For more general Lorentz and flavor structures it will be expedient to define compositeness scales 
that are directly comparable to those in~\ref{Lnew}.
For example, when limits are quoted relative to an operator basis using left- and right-chiral fields,
then both, the operators and their coefficients should be rotated so that the norm is preserved, i.e.,
\begin{equation}
\label{norm}
\sum\limits_{k,l =V,A} g_{kl}^2 = N \sum\limits_{i,j =L,R} g_{ij}^2\ ,
\end{equation}
with $N = 1$.
However, the formalism of Reference~\cite{Eichten:1983hw} involving ordinary chiral projectors 
leads to a rescaling of couplings ($N = 4$) and of the compositeness scales (by a factor of 2). 
This appears to be the reason why the LEP~2 Collaborations~\cite{Schael:2013ita} find generally 
much weaker limits for purely chiral operators.

\begin{figure}[!t]
\begin{center}
\includegraphics[width=0.93\textwidth]{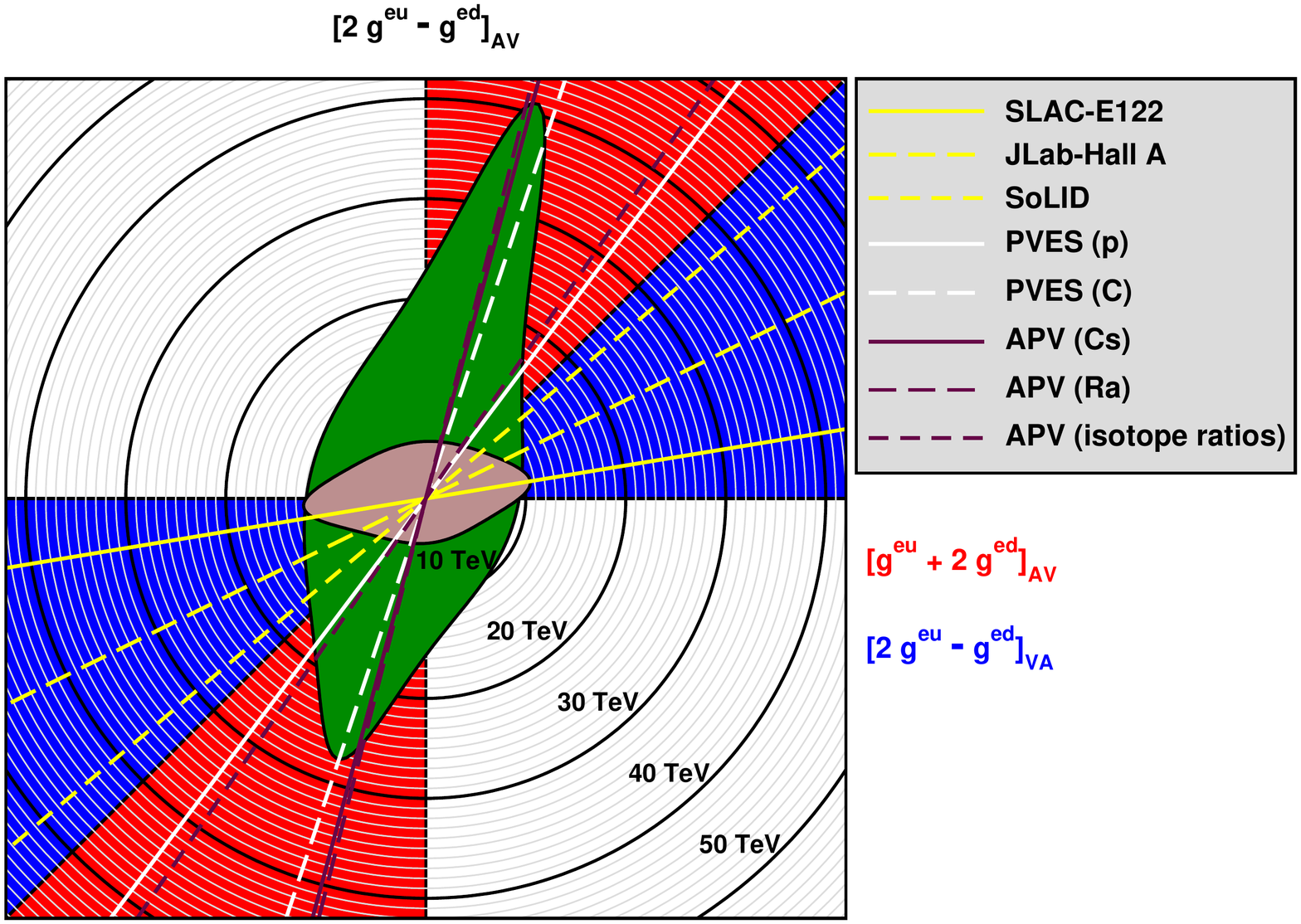}
\caption{Compositeness scales for operators in the two planes (overlaid) defined by 
$g_{AV}^{eu} + 2\, g_{AV}^{ed}$ and
$2\, g_{VA}^{eu} - g_{VA}^{ed}$ (vertical direction) vs.\
$2\, g_{AV}^{eu} - g_{AV}^{ed}$ (horizontal direction).
The blue segment is accessible to PVDIS experiments (yellow lines) and defines a plane 
(containing the brown 95\% CL exclusion contour) perpendicular to the plane 
containing the red segment, the green contour and the white and maroon lines.
Thus, the two planes are subspaces of a three-dimensional parameter space which intersect along the horizontal direction.
The lines indicate the coupling combinations of the various experiments
relative to the common horizontal direction (cf.\ the angle $\theta$ shown in {\bf Table~\ref{reach}}).
Note that we have adjusted the E122 results for sea quark dilution ($R_v \neq 1$).}
\label{scales}
\end{center}
\end{figure}

A similar complication arises if a given experiment or data set is 
sensitive to some combination of quark flavors.
As before, a rotated operator basis (without rescaling) should be used
in which one of the operators coincides with the one relevant to the observable in question.
It is the constraint on this operator, or equivalently the lower limit on the associated compositeness scale,
which concrete models of new physics ought to satisfy.
For example, elastic $ep$ scattering probes the operator,
\be
\label{udoperator2}
\left[ {2\, g_{AV}^{eu} + g_{AV}^{de} \over \sqrt{5}} \right]
\frac{\bar{e} \gamma^\mu \gamma^5 e}{2 v^2}
\left( {2\, \bar{u} \gamma_\mu u + \bar{d} \gamma_\mu d \over \sqrt{5}}\right)\ ,
\end{equation}
where the scale $\Lambda^{ep}_{AV}$ may be defined such
that the bracketed prefactor assumes the fixed reference value $4\pi$.
Thus, the constraint~\ref{QweakI} translates to
\begin{equation}
\Lambda_+ > v \sqrt{\frac{\sqrt{5}\, 8\pi}{|2\, g_{AV}^{eu} + g_{AV}^{ed}|^+}} = 15.3~{\rm TeV}\ , \qquad\qquad 
\Lambda_- >  19.0~{\rm TeV}\ .
\end{equation}
We summarize the current and expected compositeness scale limits from PVES and APV in {\bf Table~\ref{reach}}
and illustrate the scales in {\bf Figure~\ref{scales}}.


\subsection{Leptophobic $Z's$}

Leptophobic $Z's$~\cite{Babu:1996vt}, corresponding to the class of models with additional neutral gauge 
bosons ($Z'$) with negligible couplings to leptons, are particularly interesting in the context of PVES 
when they have sizable axial couplings to quarks. 
Large backgrounds from dijet production  $p\bar{p}$, $pp\to jj$ in high energy hadron colliders, 
tend to dilute bounds on  low mass leptophobic $Z's$ with $m_{Z'}\lesssim 300$~GeV.  
Planned precision measurements in electron-deuteron PVDIS  can provide strong constraints 
precisely in this region of parameter space where hadron collider constraints are weakest. 

Leptophobic $Z's$ with axial couplings to quarks typically generate sizable shifts 
in the $g^{eq}_{VA}$ coefficients while the $g^{eq}_{AV}$ are relatively unaffected. 
The latter are best measured by the Qweak, P2, and APV experiments and 
in $A_{LR}^{\rm DIS}$ in Equation~\ref{eDIS} they can be treated as known quantities. 
SoLID is expected to measure the combination $2\, g^{eu}_{VA}-g^{ed}_{VA}$ with an uncertainty 
of about $\pm 0.007$, and is thus particularly well-suited to constrain such leptophobic $Z'$ scenarios.

\begin{figure}[!t]
\begin{center}
\includegraphics[width=0.93\textwidth]{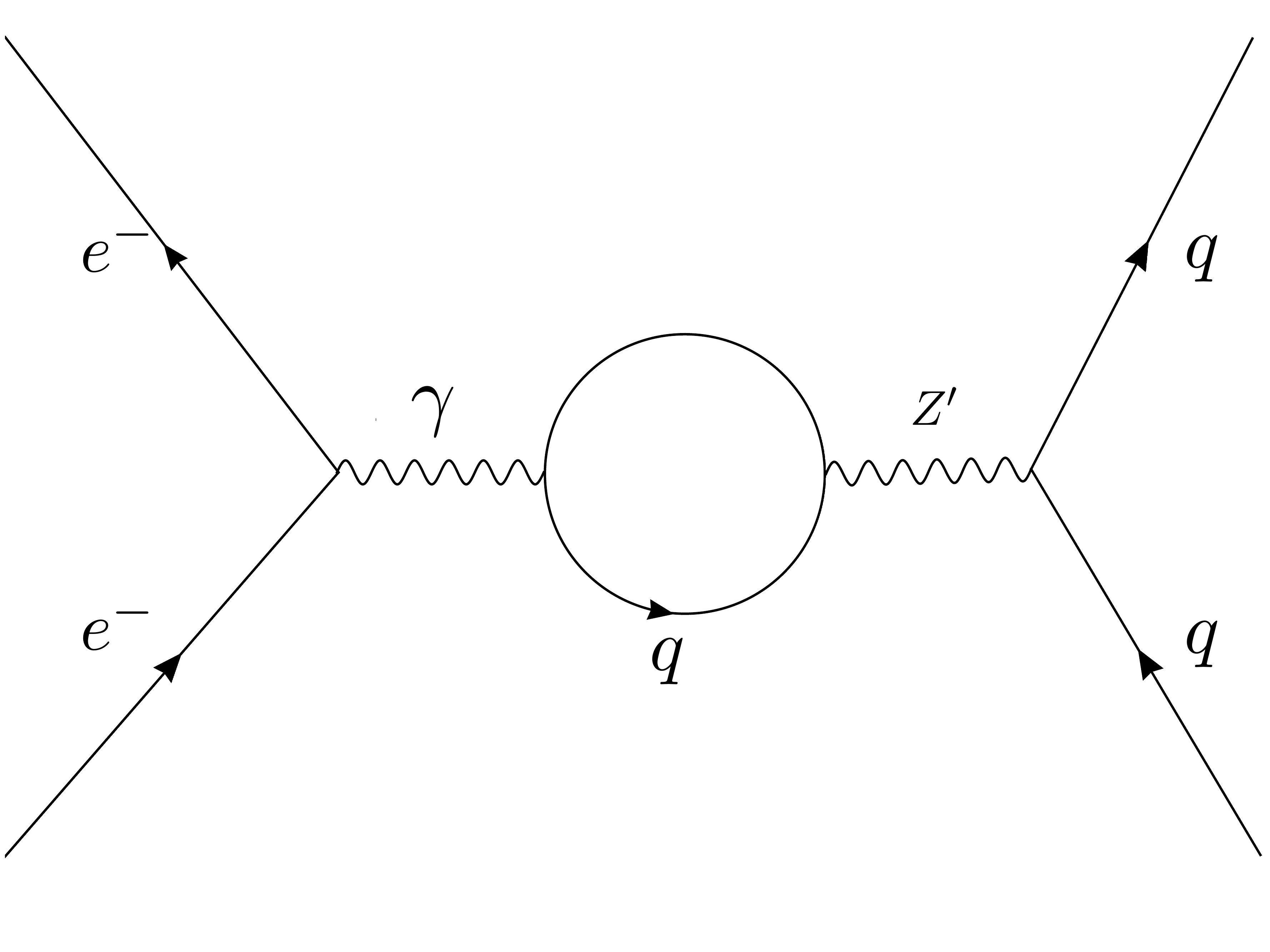}
\caption{Dominant one-loop contribution of a leptophobic $Z'$ to low energy parity violation.
Note, that the $Z'$ needs both vector and axial-vector couplings to produce this effect.}
\label{leptoZ}
\end{center}
\end{figure}

The dominant shift in the $g^{eq}_{VA}$ coefficients arise~\cite{Buckley:2012tc} from $\gamma$-$Z'$ mixing, 
as shown in {\bf Figure~\ref{leptoZ}}. 
Note that since the electron couplings are negligible, 
only quarks contribute to the $\Pi_{\gamma Z'}(q^2)$ two-point correlation function.  
In principle, $Z$-$Z'$ mixing can also contribute, 
causing shifts in both the $g^{eq}_{AV}$ and $g^{eq}_{VA}$ coefficients. 
However, the $Z$-$Z'$ mixing angle is constrained~\cite{Erler:2009jh} to be less than $10^{-2}$
and mostly affects the overall normalizations, so that this contribution is negligible. 

The leptophobic $Z'$ scenario explored in Reference~\cite{Buckley:2012tc} is expected to give rise to
$3~\sigma$ (6 to $7~\sigma$) effects in the combination $2g^{eu}_{VA}-g^{ed}_{VA}$ in the SoLID (EIC) experiments. 
This work was extended in Reference~\cite{GonzalezAlonso:2012jb} to include vertex corrections from $Z'$ loops, 
for the case where the $Z'$ only couples to the light quarks which allows a systematic expansion in $m_{q}^2/m_{Z'}^2$.   
These analyses show that a large shift in the $g^{eq}_{VA}$ coefficients, with the $g^{eq}_{AV}$ relatively unaffected, 
would be a strong indication of a leptophobic $Z'$ with axial-vector quark couplings.

\subsection{Dark $Z$}

The ``Dark $Z$" scenario~\cite{Davoudiasl:2012ag,Davoudiasl:2012qa,Davoudiasl:2012ig} involves 
a light vector boson $Z_d$, with a mass in the range $10~{\rm MeV} < m_{Z_d} < 10~{\rm GeV}$, 
arising from a new spontaneously broken $U(1)_d$ gauge symmetry associated with a hidden sector 
which may or may not be related to dark matter. 
In contrast to other $Z'$ scenarios, the physics of the $Z_d$ cannot be integrated out 
and absorbed into contact interactions as long as $m_{Z_d}^2 \lesssim Q^2$; 
instead it has to be treated as a low mass dynamical degree of freedom. 
This is particularly interesting for low energy parity violation since 
there is an enhancement of the $Z_d$ propagator relative to the $Z$ propagator.

Before spontaneous symmetry breaking, the kinetic mixing term,
\be
{\cal L}_{\rm kin.} = \frac{\varepsilon}{2\cos \theta_W} B_{\mu \nu} Z_{d}^{\mu \nu}\ ,
\ee
gives rise to $\gamma$-$Z_d$ and $Z$-$Z_d$ mixing. 
The mixing parameter is constrained~\cite{Batell:2009yf,Bjorken:2009mm,Jaeckel:2010ni} to be 
$\varepsilon \lesssim 10^{-3}$ from studies of the so-called dark photon scenario~\cite{Fayet:1980ad,Fayet:1980ss,Fayet:1981rp}, 
where charged particles couple to the $Z_d$ by generating mixing with the photon. 

More recently~\cite{Davoudiasl:2012ag,Davoudiasl:2012qa,Davoudiasl:2012ig}, the effects of the $Z$-$Z_d$ 
mass mixing matrix were investigated. 
After spontaneous symmetry breaking, it has the general form
\be
M_0 = m_Z^2 
\left( \begin{array}{cc} 1 & -\varepsilon_Z  \\ - \varepsilon_Z & m_{Z_d}^2/m_Z^2 \end{array} \right)\ , \qquad\qquad 
\varepsilon_Z = \frac{m_{Z_d}}{m_Z}\, \delta\ ,
\ee
where  $\delta$ is a model-dependent quantity and the contribution from $\varepsilon$ 
appears at ${\cal O}(\varepsilon^2)$ and is thus negligible. 
The constraint $0\leq\delta^2 <1$ is needed to avoid an infinite-range or tachyonic $Z_d$. 
The $Z_d$ couples to the EM and weak neutral currents through,
\be
\label{darkZint}
{\cal L}_{\rm int.} = - {g \over \sqrt{2}} \left[ \varepsilon J^\mu_A + \frac{\varepsilon_Z}{2\cos \theta_W}\, J^\mu_Z \right] {Z_d}_\mu\ .
\ee
The net effect of the new interactions in Equation~\ref{darkZint} for PV amplitudes of the form
${\cal M}^{PV}_{\rm NC} = v^{-2} F(\sin^2\theta_W)/4$ is obtained through the replacements
\be
{1\over v^2} \to {\rho_d \over v^2}\ , \qquad\qquad
\sin^2\theta_W \to \kappa_d\> \sin^2\theta_W\ ,
\ee
where the quantities $\rho_d$ and $\kappa_d$ are given by,
\be
\rho_d = 1+ \delta^2 \frac{m_{Z_d}^2}{Q^2 + m_{Z_d}^2}\ , \qquad\qquad
\kappa_d = 1 - \varepsilon\, \frac{m_Z}{m_{Z_d}}\, \delta\, \frac{\cos\theta_W}{\sin\theta_W}\, \frac{m_{Z_d}^2}{Q^2 + m_{Z_d}^2}\, .
\ee
As long as there are no accidental cancellations between the effects of $\rho_d$ and $\kappa_d$, 
the strongest bound on the $Z_d$ scenario, over its entire mass range, typically comes from APV. 
The shift in the Cs weak charge, $Q_W^{\rm Cs}$, from its SM value is~\cite{Davoudiasl:2012ag},
\be
\Delta Q_W^{\rm Cs} \approx \delta^2 \left[ Q_W^{\rm Cs}({\rm SM}) 
+ 220\, \frac{\varepsilon}{\varepsilon_Z} \cos \theta_W \sin \theta_W \right]\ .
\ee
For $\varepsilon \ll \varepsilon_Z$ one obtains the 90\% CL bound, $\delta^2\lesssim 0.006$. 
On the other hand, for $\varepsilon \simeq \varepsilon_Z $ a cancellation between the two terms above can dilute the bound. 
Independent but similar bounds~\cite{Davoudiasl:2012ag} arise from 
M\o ller scattering~\cite{Czarnecki:2000ic,Anthony:2005pm}, primarily through its constraint on $\kappa_d$, 
for $\varepsilon \simeq \varepsilon_Z$ and $m_{Z_d}^2\gg Q^2 \simeq (0.16\> {\rm GeV})^2$. 
For very light masses, $m_{Z_d}\lesssim 200~{\rm MeV}$, the bound from M\o ller scattering 
is weaker than that from APV. 
However, there do exist regions in parameter space where M\o ller scattering can provide stronger bounds. 
E.g., for $\varepsilon \simeq 2\times 10^{-3}$ and $m_{Z_d}\simeq 100~{\rm MeV}$, 
corresponding to the favored region to explain the observed discrepancy~\cite{Bennett:2006fi,Fayet:2007ua,Pospelov:2008zw} 
in the muon anomalous magnetic moment, M\o ller scattering gives the bound $|\delta| < 0.01$. 
Ongoing and proposed experiments at JLAB~\cite{McKeown:2011yj,Androic:2013rhu} 
and in Mainz~\cite{Becker:2013fya} are expected to improve the bound by an order of magnitude.

Complementary constraints~\cite{Davoudiasl:2012ag} arise from rare $K$ and $B$ meson decays, 
$K\to \pi Z_d$ and $B\to K Z_d$, respectively, through flavor changing neutral currents mediated by the $Z_d$. 
The suppression factor  $m_{Z_d}/m_{Z}$, from the induced coupling to the $Z_d$ is overcome 
by the enhancement factors $m_{K}/m_{Z_d}$ and $m_{B}/m_{Z_d}$ in the longitudinally polarized $Z_d$ channel. 
If the lifetime of the $Z_d$ is long enough or if it decays as $Z_d\to \nu \bar{\nu}$ or to hidden sector particles, 
it will appear as a missing energy signal in these rare decays. 
Otherwise it can decay to $Z_d\to \ell^+\ell^-$ and appear as a displaced vertex. 
The pattern of $Z_d$ decays is, of course, model dependent. 
The $K$-meson decays lead to constraints of the form $|\delta | \lesssim 0.01/\sqrt{{\rm BR}(Z_d\to e^+e^-)}$ 
and $|\delta | \lesssim 0.001/\sqrt{{\rm BR}(Z_d\to {\rm missing \>energy})}$, 
where the overall numerical factor is model dependent. 
Similarly, the $B$-meson decays lead constraints of the form $|\delta | \lesssim 0.001/\sqrt{{\rm BR}(Z_d\to e^+e^-)}$ 
and $|\delta | \lesssim 0.01/\sqrt{{\rm BR}(Z_d\to {\rm missing \>energy})}$. 

Precision studies of Higgs decays at the LHC can also yield useful bounds~\cite{Davoudiasl:2012ag} on the $Z_d$ scenario. 
In particular, the $H\to ZZ_d$ mode can be constrained through studies of the $H\to ZZ^*\to \ell^+\ell^-\ell^+\ell^-$ and
$H\to ZZ^*\to \ell^+\ell^-\nu\bar{\nu}$ channels.  
Just as for the rare $K$ and $B$ meson decays, the suppression factor of $m_{Z_d}/m_Z$ from the induced coupling to 
the $Z_d$ is overcome by an enhancement factor $m_{H}/m_{Z_d}$ in the longitudinally polarized $Z_d$ channel. 
For $m_H=125$~GeV one typically finds the constraint $\Gamma(H\to ZZ_d/\Gamma_H^{\rm SM})\simeq 16\, \delta^2 \lesssim 0.1$. 
Eventually, with enough statistics, precision Higgs studies could yield useful independent bounds on $\delta$.

Thus, low energy PV experiments can be used in concert with constraints 
coming from rare $K$, $B$ and Higgs decays to study the $Z_d$ scenario.

\section{CONCLUSIONS}
\label{secconclusion}


Many PVES experiments have been completed or are in various stages of progress.
These experiments have a large physics reach, both in terms of the strong and weak interactions.  
They can search for physics beyond the standard model with little theoretical uncertainty.  
In terms of a hypothetical compositeness scale, PVES experiments are among the most sensitive.  
They can also probe the dark sector at a low mass level, and search for a leptophobic Z' of intermediate mass.

PVES also provides a unique window on hadronic structure.  
For example, since the weak charge of the neutron is large, PVES experiments probe 
the radius of the density of neutrons in nuclei like $^{208}$Pb and $^{48}$Ca.  
In turn, this provides critical information for interpreting the physics of neutron stars.  
PVES can also study CSV at the parton level and provide a unique window to HT terms in DIS.

\section*{Acknowledgment} 
We thank Krishna Kumar for useful comments on the manuscript. J.E. acknowledges support by PAPIIT (DGAPA--UNAM) project IN106913, by CONACyT (M\'exico) project 151234,
and by the Mainz Institute for Theore-tical Physics (MITP).
P.A.S. is funded in part by the Department of Energy (DOE) under grant DE-FG02-84ER40146.
C.H. acknowledges support by DOE grants DE-FG02-87ER40365 (Indiana University) 
and DE-SC0008808 (NUCLEI SciDAC Collaboration). S.M. is supported by Northwestern University.

\bibliographystyle{arnuke_revised2}
\bibliography{review}
\end{document}